\documentclass[fullpage]{article}
\usepackage[affil-it]{authblk} 
\usepackage{etoolbox}
\usepackage{lmodern}
\usepackage[a4paper]{geometry}
\usepackage[utf8]{inputenc}
\usepackage{amsmath}
\usepackage{graphicx}
\usepackage[dvipsnames]{xcolor}
\usepackage[colorlinks]{hyperref}
\hypersetup{linkcolor=blue,citecolor=blue,urlcolor=blue}

\newcommand{\Odd}{O(d,d)}

\newcommand{\ap}{\alpha'}

\makeatletter
\patchcmd{\@maketitle}{\LARGE \@title}{\fontsize{16}{19.2}\selectfont\@title}{}{}
\makeatother

\title{O$(d,d)$ covariant String Cosmology to all orders in $\alpha^{\prime}$}

\author[1]{Heliudson Bernardo\footnote{Email: \href{mailto:heliudson@hep.physics.mcgill.ca}{heliudson@hep.physics.mcgill.ca}}}
\author[2]{Robert Brandenberger\footnote{Email: \href{mailto:rhb@hep.physics.mcgill.ca}{rhb@hep.physics.mcgill.ca}}}
\author[3]{Guilherme Franzmann\footnote{Email: \href{mailto:guilherme.franzmann@su.se}{guilherme.franzmann@su.se}}}

\affil[1,2]{Physics Department, McGill University, Montreal, QC, H3A 2T8, Canada}
\affil[3]{The Oskar Klein Centre, Department of Physics, Stockholm University, Stockholm 106 91, Sweden}
\affil[3]{Nordita and KTH Royal Institute of Technology, Roslagstullsbacken 23, SE-106 91 Stockholm, Sweden}

\date{}

\begin{document}

\maketitle

\begin{abstract}

Recently, all duality invariant $\alpha'$-corrections to the massless NS-NS sector of string theory on time-dependent backgrounds were classified and the form of their contribution to the action were calculated. In this paper we introduce matter sources in the resulting equations of motion in an O$(d,d)$ covariant way. We show that either starting with the corrected equations and sourcing them with matter or considering corrections to the matter sourced lowest order equations give the same set of equations that defines string cosmology to all orders in $\alpha'$. We also discuss perturbative and non-perturbative de Sitter solutions including matter.
 
\end{abstract}

\tableofcontents

\section{Introduction}

String theory is the main candidate for a unified theory of quantum gravity. Despite of many advances in understanding its non-perturbative structure, most of the calculations are still done in perturbation theory. There are two expansion parameters in string theory: the dimensionless string coupling constant, $g_s$, and the dimensionful string length, $\sqrt{\alpha'}$ \cite{Polchinski:1998rq, Polchinski:1998rr}. While expansion in the former is basically the $\hbar$ expansion present also in particle-like theories, the expansion in $\alpha'$ is purely stringy and can be thought of as coming from corrections due to the extended nature of the strings. Therefore, low-energy effective supergravity theories at energy scales much smaller than $1/\sqrt{\alpha'}$, which implies curvature scales $\mathcal{R}$ small compared with the string length, are approximations to the full spacetime theory, that  should include all $\alpha'$ contributions. Thus, these theories defined in the regime of large curvature, $\mathcal{R}/\sqrt{\ap} \gg 1$, cannot be trusted as we approach the string scale, and they must be corrected non-perturbatively at such scales.

Given that supergravities are low-energy theories for the massless spectrum of the strings, which includes the metric, these $\alpha'$-corrections are important for having a sensible description of the fields in regions close to apparent singularities which arise in the perturbative description, possibly removing them even in the absence of quantum effects. Thus, once we are able to include $\ap$-corrections in a consistent manner for cosmological backgrounds, we can study whether singular solutions can be smoothed out. Note that cosmology offers a very promising window to explore observational consequences of string theory. Thus, studying cosmological solutions in string theory is a very well motivated path to pursue \cite{Brandenberger:1988aj,Tseytlin:1991xk} (for reviews on String Cosmology see \cite{Erdmenger:2009zz, Baumann:2014nda,Gasperini:2007zz}, and references therein). Also, since computing stringy effects on arbitrary backgrounds is a very laborious task, time-dependent but space-independent settings may help to understand corrections for simple but non-trivial backgrounds.

Indeed, early studies \cite{Meissner:1991zj, Meissner:1991ge} in String Cosmology showed that under a cosmological  ansatz the $D=d+1$ dimensional supergravity action has a global O$(d,d)$ symmetry, generalizing the previously known scale factor duality \cite{Veneziano:1991ek}. In \cite{Gasperini:1991ak}, an explicitly O$(d,d)$ covariant formalism was developed, with a manifestly duality invariant action. Moreover, in \cite{Sen:1991zi} it was shown that the O$(d,d)$ symmetry is present to all orders in $\alpha'$ provided no fields depend on the $d$ spatial coordinates. On the other hand, the first order $\alpha'$-corrections to supergravity were computed in the late 1980's \cite{Metsaev:1987zx, Hull:1987pc, Gross:1986mw}, and in past years the use of dualities have helped the calculation of corrections in Double Field Theory (DFT) (see \cite{Siegel:1993th, Hull:2009mi} for original articles on DFT and \cite{Hohm:2013bwa, Aldazabal:2013sca} for reviews). For $\alpha'$-corrections from DFT see \cite{Hohm:2013jaa, Hohm:2014xsa,Marques:2015vua, Baron:2017dvb}, and references therein. 

Recently, motivated by the impact which winding modes have in toroidal compactifications, cosmological solutions in Double Field Theory started to be explored \cite{Wu:2013sha, Brandenberger:2018bdc, Bernardo:2019pnq, Angus:2018mep, Angus:2019bqs}. One of the aims of these studies was to explore the nature of the solutions in situations where singularities in the supergravity approximation arise. However, no $\alpha'$-corrections were considered, and such corrections are expected to be important.  Thus, any development towards finding $\alpha'$-corrections to Double Field Theory will have important consequences for these solutions. It is worth stressing that the duality coming from toroidal backgrounds is actually relating two seemly different but physically equivalent backgrounds, while the global O$(d,d)$ duality discussed in \cite{Gasperini:1991ak}, and in this paper, relates physically different backgrounds. In the latter case, given a solution we can construct a different new one by applying an O$(d,d)$ transformation to the first, while T-duality relates physically equivalent backgrounds. From now on, we use the term "duality" to refer to the global O$(d,d)$, that may not be related with T-duality since the space could be compact or not.

Although the O$(d,d)$ transformations also receive $\alpha'$-corrections, it was shown in \cite{Meissner:1996sa} that we can redefine the fields such that the form of the transformations are preserved, at least to first-order in $\alpha'$. Assuming that this is the case to all orders, i.e., that there are field variables in terms of which the duality transformations are preserved, it was shown in \cite{Hohm:2019jgu} that it is possible to classify all O$(d,d)$ invariant $\alpha'$-corrections. Hohm and Zwiebach considered a formalism with manifest O$(d,d)$ symmetry and showed that there are field variables which allow the full action to be compactly written in terms of O$(d,d)$ covariant fields. 

In spite of \cite{Hohm:2019jgu} having established significant progress in understanding the $\alpha'$-corrections on cosmological backgrounds, only the vacuum theory was developed and no matter sources were considered. The conceptual leap from General Relativity to Cosmology requires the introduction of matter and energy, ordinarily through the energy momentum tensor of a fluid. Hence, in order to have not only a time-dependent background in string theory, but also a framework to study Cosmology in string theory including the $\alpha'$-corrections, we need the gravitational sector to be coupled to matter. The matter sources can be introduced in the context of string theory or purely phenomenologically. In the former approach we could study the massive modes of the strings, given by an infinite tower of excited states. This has been modelled as a gas of strings \cite{Brandenberger:1988aj, Tseytlin:1991xk, Gasperini:1991ak}. In the latter approach we simply postulate a phenomenological energy momentum tensor that is used to construct stringy inspired models. In fact, the inclusion of matter to the framework was recognized as a future direction in the conclusion of \cite{Hohm:2019jgu} . 

The goal of this paper is to push forward this program and construct a full $\alpha'$-corrected manifestly O$(d,d)$ covariant formulation of String Cosmology in the presence of matter. At lowest order in $\alpha'$, this was partially accomplished in \cite{Gasperini:1991ak} (see also \cite{Gasperini:2007zz}). We base our analysis of the matter sector on \cite{Gasperini:1991ak}, and use field redefinitions inspired by the ones in \cite{Hohm:2019jgu} to write the String Cosmology equations in a duality covariant way. We derive a set of equations that can be used to explore the $\alpha'$-corrections to several possible setups in String Cosmology, for instance the String Gas Cosmology scenario \cite{Brandenberger:1988aj, Brandenberger:2008nx, Battefeld:2005av}, Pre-Big Bang cosmology \cite{Gasperini:2002bn} and the more recent string black-hole gas picture \cite{Quintin:2018loc}. We also find perturbative solutions and discuss conditions for de Sitter solutions non-perturbatively, including matter fields. 

The paper is organized as follows. In Section \ref{sec:lowest_order_vacuum}, we provide a brief review of the results obtained in \cite{Gasperini:1991ak}, highlighting how to get an O$(d,d)$ covariant description of fluids. In Section \ref{sec:matter_coupling}, we present a summary of the results from \cite{Hohm:2019jgu}. In Section \ref{sec:alpha_corrected_matter}, we show the duality covariant equations for String Cosmology, proving that we get the same structure for such equations regardless of including matter before or after field redefinitions.  Section \ref{sec:alpha_cosmo} introduces the $\alpha'$-corrected Friedmann equations while Section \ref{sec:solutions} provides a discussion of perturbative and de Sitter solutions. In Section \ref{sec:conclusion} we conclude.
 
\section{Lowest order vacuum action and $\alpha'$-corrections} \label{sec:lowest_order_vacuum}

The gravitational sector of all superstring theories includes the fields coming from the massless level of the  bosonic string theory: the metric, $G_{\mu\nu}(x)$, the Kalb-Ramond 2-form field $B_{\mu\nu}(x)$ and the dilaton $\phi(x)$. The low-energy effective action for these fields is given by
\begin{equation}\label{sugraaction}
    S_0 = \frac{1}{2\kappa^2}\int d^D x \sqrt{-G} e^{-2\phi}\left(R + 4 G^{\mu\nu}\partial_\mu \phi \partial_\nu \phi -\frac{1}{12}H_{\mu\nu\rho}H^{\mu\nu\rho}\right),
\end{equation}
where $H_{\mu\nu\rho}$ is the field strength for $B_{\mu\nu}$ and $R$ is the metric's Ricci scalar. For a cosmological background the ansatz for the fields is given as $G_{00} = -n^2(t)$, $G_{0i} = 0$, $G_{ij} = g_{ij}(t)$, $B_{00} = 0$, $B_{0i} = 0$, $B_{ij} = b_{ij}(t)$ and $\phi = \phi(t)$, and the action can be written as \cite{Meissner:1991zj,Gasperini:2007zz}
\begin{equation}\label{oddaction}
    S_0 = \int d^dx I_0, \quad I_0 = -\frac{1}{2\kappa^2}\int dt n e^{-\Phi}\left[(\mathcal{D}\Phi)^2 + \frac{1}{8}\text{tr}(\mathcal{DS})^2\right],
\end{equation}
where $\mathcal{D} = 1/n \partial_t$, $\Phi = 2\phi - \ln \sqrt{\det g}$ and $D = d+1$. The trace is on the $2d$ indices of the matrix
\begin{equation}
    \mathcal{S} = \eta \mathcal{H} = \begin{pmatrix}
    bg^{-1} & g- bg^{-1}b \\
    g^{-1} & - g^{-1}b 
    \end{pmatrix},
    \end{equation}
where
\begin{equation}
    \eta  = \begin{pmatrix}
    0 & 1 \\
    1 & 0 
    \end{pmatrix}, \quad 
    \mathcal{H} = \begin{pmatrix} 
    g^{-1} & -g^{-1}b \\ 
    bg^{-1} & g - bg^{-1}b
    \end{pmatrix},
\end{equation}
with $\eta$ being an $\Odd$ metric and $\mathcal{H}\in\Odd$, such that $\mathcal{S}^2 = 1$, while $\Phi$, the shifted dilaton, is a scalar under $\Odd$ transformations. In the form (\ref{oddaction}), the action $I_0$ is manifestly invariant under the global $\Odd$ transformations and time reparameterizations. The equations of motion are \cite{Gasperini:1991ak, Gasperini:2007zz, Hohm:2019jgu}
\begin{align}
    E_{\Phi} = \frac{1}{2\kappa^2}\left(2\mathcal{D}^2\Phi - (\mathcal{D}\Phi)^2 +\frac{1}{8} \text{tr}(\mathcal{DS})^2\right) &= 0 \quad (\Phi)\\
    E_{\mathcal{S}}=\frac{1}{8\kappa^2}\left(\mathcal{D}^2\mathcal{S} + \mathcal{S}(\mathcal{DS})^2 - (\mathcal{D}\Phi)(\mathcal{DS})\right) &= 0 \quad (\mathcal{S})\\
    E_n = \frac{1}{2\kappa^2}\left((\mathcal{D}\Phi)^2 + \frac{1}{8}\text{tr}(\mathcal{DS})^2\right) &= 0 \quad (n).
\end{align}
Instead of the equation for $\mathcal{S}$, we can use the conservation of the $\Odd$ charge to close the system \cite{Hohm:2019jgu},
\begin{equation}
    \frac{d\mathcal{Q}}{dt} = - 2e^{-\Phi}\mathcal{S}E_{\mathcal{S}} = 0,
\end{equation}
where
\begin{equation}
    \mathcal{Q} \equiv -\frac{1}{2}e^{-\Phi}\mathcal{S}\dot{\mathcal{S}}.
\end{equation}

In \cite{Hohm:2019jgu}, it is shown that one can use field redefinitions to bring corrections of order $\alpha'^k$ into the form
\begin{equation}
    S_k = \frac{1}{2\kappa^2}\int d^d x I_k, \qquad I_k = \alpha'^k \int dt n e^{-\Phi} X(\mathcal{DS}), \label{eq:action_vacuum}
\end{equation}
 with $X$ being a strictly arbitrary function of $\mathcal{DS}$, with no extra time derivatives appearing\footnote{We use $\{A\}$ in the argument of functions to denote a general dependence, i.e., a function $F(\{A\})$ depends on $A$ and on its derivatives.} and no factors of $(\mathcal{DS})^2$. To prove this, it was assumed that the $(k-1)$th order action has the desired form, and then it was shown that by doing field redefinitions at order $k$ in $\alpha^{\prime}$, one can write the $k$th order action as (\ref{eq:action_vacuum}). Crucial was the fact that only the $0$th order action should be varied under the field redefinitions; the higher order terms contribute to the $\mathcal{O}(\alpha'^{k+1})$ action only. Explicitly, under the transformation
 \begin{equation}
     \Phi \rightarrow \Phi + \alpha'^k\delta \Phi, \qquad \mathcal{S} \rightarrow \mathcal{S} + \alpha'^k \delta \mathcal{S} \label{eq:field_redef}
 \end{equation}
 the variation of the second term in
\begin{equation}
    I = I_0 + \sum_{p=1}^{k-1} I_p + \alpha'^k \int dt n e^{-\Phi} X(\{\mathcal{D}\Phi\}, \{\mathcal{S}\}) + \mathcal{O}(\alpha'^{k+1}),
\end{equation}
contributes to $\mathcal{O}(\alpha'^{k+1})$ only. The proof was achieved by showing the following properties:
\begin{enumerate}
    \item A factor $\mathcal{D}^2\Phi$ in the action can be replaced by a factor
    \begin{equation}
        Q_{\Phi} = \frac{1}{2}(\mathcal{D}\Phi)^2 - \frac{1}{16}\text{tr}(\mathcal{DS})^2, \label{eq:property_1}
    \end{equation}
    that contains only first order derivatives of $\Phi$;
    \item A factor $\mathcal{D}^2\mathcal{S}$ in the action can be replaced by a factor
    \begin{equation}
        Q_{\mathcal{S}} = -\mathcal{S}(\mathcal{DS})^2 + \mathcal{D}\Phi\mathcal{DS}, \label{eq:property_2}
    \end{equation}
    that contains only first order derivatives of $\mathcal{S}$;
    \item Any action can be reduced so that it only contains first time derivatives of $\Phi$ and $\mathcal{S}$. This is based on integrating by parts the higher order derivatives until only second order derivatives are left, and then using the previous properties to get rid of them. Formally, this is equivalent to changing the derivatives $\mathcal{D}$ in terms of the form
    \begin{align}
        \alpha'^k \int dt n e^{\Phi} X({\mathcal{D}\Phi}, \{\mathcal{S}\})\mathcal{D}^{p+2}\Phi, \quad         \alpha'^k \int dt n \text{tr}(\mathcal{F}\mathcal{D}^{p+2}\mathcal{S}),
    \end{align}
    to $\Bar{\mathcal{D}}$, which acts on a function $F$ of $\mathcal{S}$, $\mathcal{DS}$ and $\mathcal{D}\Phi$ as
    \begin{equation}
        \Bar{\mathcal{D}}F = \left. \mathcal{D}F \right \vert_{\mathcal{D}^2\Phi \rightarrow Q_{\Phi}, \mathcal{D}^2\mathcal{S}\rightarrow Q_{\mathcal{S}}}.
    \end{equation}    
    \item Any action of $k$th order in $\alpha^{\prime}$ is equivalent to one without any appearance of $\mathcal{D}\Phi$ (for $k>0$). This is shown by using field redefinitions of $\Phi$; 
    \item  Any term containing a factor of $\text{tr}(\mathcal{D}\mathcal{S})^2$ in the action for the corrections can be removed by redefining the lapse function $n(t)$.
\end{enumerate}

Notice that all these properties follow from the structure of the variations of the action with respect to the fields, i.e., they follow from the forms of the $E_{\Phi}$, $E_{\mathcal{S}}$, $E_{n}$, which are related to the forms of $Q_{\Phi}$ and $Q_{\mathcal{S}}$.

For a Friedmann-Lema\^itre-Robertson-Walker (FLRW)  ansatz, $g_{ij} = a^2(t)\delta_{ij}$ and $b_{ij} = 0$, it was also shown in \cite{Hohm:2019jgu} that only the single trace higher order terms contribute and the corrected action to all orders in $\alpha^{\prime}$ is given by
\begin{equation}
    I = \frac{1}{2\kappa^2}\int dt n e^{\Phi} \left(- (\mathcal{D}\Phi)^2 +\sum_{k=1}^{\infty}\alpha'^{k-1}c_k \text{tr}(\mathcal{DS})^{2k}\right), \quad c_1= -\frac{1}{8}, \label{eq:ActionCorrected}
\end{equation}
where the values of the coefficients $\{c_k\}$ for $k\neq1$ are not known\footnote{Determining the values of the coefficients would require more input from string theory - not just symmetry consideration.} and take different values for different string theories ($c_2 = 1/64$ for heterotic strings and $c_2=0$ for type II strings, for instance). Note that (\ref{eq:ActionCorrected}) is a large simplification over what could be expected from higher-derivative corrections for (\ref{sugraaction}): not only does the dilaton contribution have the same structure, but there are no extra time derivatives of $\mathcal{D S}$.

The $\ap$-corrected equations of motion are then (setting $n=1$)
\begin{align}\label{vacuumequations}
    2\kappa^2E_{\Phi} = 2\Ddot{\Phi} - \dot{\Phi}^2 - \sum_{k=1}^{\infty}\alpha'^{k-1}c_k \text{tr}\dot{\mathcal{S}}^{2k} &= 0 \\
    2\kappa^2E_n = \dot{\Phi}^2 - \sum_{k=1}^{\infty}\alpha'^{k-1}(2k-1)c_k \text{tr}\dot{\mathcal{S}}^{2k} &= 0\label{vaccumequations2}\\
    \frac{d}{dt}\mathcal{Q} \equiv \frac{d}{dt}\left(e^{-\Phi}\sum_{k=1}^{\infty}\alpha'^{k-1}4kc_k\mathcal{S}\dot{\mathcal{S}}^{2k-1}\right) &= 0,\label{vacuumequations3}
\end{align}
where we stressed that the equation of motion for $\mathcal{S}$ is equivalent to the conservation of the O$(d,d)$ charge $\mathcal{Q}$. Using $g_{ij} = a^2(t) \delta_{ij}$ inside $\mathcal{S}$, these questions simplify to \cite{Hohm:2019jgu}
\begin{subequations}\label{HZequations}
    \begin{equation}
     \dot{\Phi}^2 + HF'(H) - F(H) = 0
    \end{equation}
    \begin{equation}
     \frac{d}{dt}F'(H) - \dot{\Phi}F'(H) = 0
    \end{equation}
    \begin{equation}
      2 \Ddot{\Phi} - \dot{\Phi}^2 + F(H) = 0,  
    \end{equation}
\end{subequations}
where 
\begin{equation}
    F(H) \equiv 2d \sum_{k=1}^{\infty}(-\alpha')^{k-1}c_k 2^{2k} H^{2k},
\end{equation}
and $'$ denotes a derivative with respect to the Hubble parameter, $H\equiv \dot{a}/a$. 

It was shown in \cite{Hohm:2019ccp, Hohm:2019jgu} that under some assumptions on the function $F(H)$ there are de Sitter solutions to the equations (\ref{HZequations}), and that we can construct perturbative solutions by solving the equations order by order\footnote{ It is possible (in light of the {\it swampland} conditions (see \cite{Brennan:2017rbf, Palti:2019pca} for reviews) on effective field theories which are consistent with string theory) that string theory prevents the function $F(H)$ from having the form required to admit de Sitter solutions.}. Conditions for having a de Sitter solution in the Einstein frame were studied in \cite{Krishnan:2019mkv}. Furthermore, in \cite{Wang:2019mwi} Anti-de Sitter solutions for the bosonic case were explored, and in \cite{Wang:2019kez, Wang:2019dcj} non-singular solutions were discussed. However, we should note that despite the fact that these equations define consistent time-dependent string backgrounds (once we specify the values of $c_k$), they remain pure vacuum equations for the background fields, which means there is neither energy density nor pressure sourcing them. In order to improve this framework with a consistent matter coupling, we can first consider how duality invariant stringy cosmological equations have been introduced in the presence of matter \textit{at lowest order in $\ap$} in \cite{Gasperini:1991ak}, as done in the following section.

\section{Matter coupling to the lowest order vacuum action}\label{sec:matter_coupling}

In \cite{Gasperini:1991ak} the equations of motion from the quadratic action $I_0$ were coupled with a gas of non-interacting strings, in a fully O$(d,d)$ covariant way (see also \cite{Gasperini:2007zz}). In this section we review the approach of \cite{Gasperini:1991ak} considering a slight generalization after starting from a general matter action, instead of only a gas of strings, assumed to be duality invariant. The total action is, therefore, given by
\begin{equation}
    S_T = -\frac{1}{2\kappa^2}\int d^d x dt n e^{-\Phi}\left[(\mathcal{D}\Phi)^2 + \frac{1}{8}\text{tr}(\mathcal{DS})^2\right] + S_m[\Phi, \mathcal{S}, n, \chi], \label{eq:AlphaCosmo_matter}
\end{equation}
where we generically denote the matter fields as $\chi(t)$ (with only time dependence to be consistent with the symmetries of the background). We assume that $S_m$ is $\Odd$ invariant, which is certainly true for a gas of strings \cite{Gasperini:2007zz} (see also \cite{Tseytlin:1991xk}), but it is also a reasonable assumption for semi-phenomenological analyses. We also include a dependence on the generalized dilaton field $\Phi(t)$ in the matter action in order to be as general as possible (see for instance \cite{Angus:2019bqs,Quintin:2018loc}). 

Varying the action with respect to $\Phi$ gives
\begin{subequations}
    \begin{equation}
         \delta_{\Phi} S_T = \int d^d x dt n e^{-\Phi}\delta \Phi E_{\Phi}^{T}
    \end{equation}
   \begin{equation}
        E^T_{\Phi} \equiv \frac{1}{2\kappa^2}\left[2\mathcal{D}^2\Phi - (\mathcal{D}\Phi)^2 +\frac{1}{8}\text{tr}(\mathcal{DS})^2\right] + \frac{e^{\Phi}}{n}\frac{\delta S_m}{\delta \Phi},
   \end{equation}
\end{subequations}
 and so the equation of motion for $\Phi$ is $E_\Phi^T = 0$,
 \begin{equation}
     2\mathcal{D}^2\Phi - (\mathcal{D}\Phi)^2 +\frac{1}{8}\text{tr}(\mathcal{DS})^2 = -2\kappa^2 \frac{e^{\Phi}}{n}\frac{\delta S_m}{\delta \Phi} \equiv \kappa^2 e^{\Phi} \bar{\sigma}, \label{eq:EOM_Phi}
 \end{equation}
where we defined a dilatonic charge
\begin{equation}
    \sigma \equiv -\frac{2}{\sqrt{-G}}\frac{\delta S_m}{\delta \Phi},
\end{equation}
and the bar denotes multiplication by $\sqrt{g}$, $\bar{\sigma} \equiv \sqrt{g}\sigma$.

Varying with respect to $\mathcal{S}$ gives  \cite{Gasperini:1991ak}
\begin{subequations} \label{variationS}
    \begin{equation}
         \delta_{\mathcal{S}} S_T = \int d^d x dt n e^{-\Phi}\text{tr}(\delta \mathcal{S} F_{\mathcal{S}})
    \end{equation}
      \begin{equation}
        F_{\mathcal{S}} \equiv -\frac{1}{8\kappa^2}(\mathcal{D}\Phi\mathcal{DS}-\mathcal{D}^2\mathcal{S})+ \frac{e^{\Phi}}{2 n} \frac{\delta S_m}{\delta \mathcal{S}}.
    \end{equation}
\end{subequations}
As explained in \cite{Hohm:2019jgu}, due to the fact that $\mathcal{S}^2 = 1$ we need to vary the action in terms of an unconstrained variable, such that the equations of motion are not simply the vanishing of $F_{\mathcal{S}}$. In fact, defining the unconstrained variation $\delta K$ as
\begin{equation}
    \delta \mathcal{S} =\frac{1}{2}(\delta K - \mathcal{S}\delta K\mathcal{S}),
\end{equation}
the equations of motion follow from
\begin{equation}
    E_{\mathcal{S}}^T = \frac{1}{2}(F_\mathcal{S}- \mathcal{S}F_\mathcal{S}\mathcal{S}) = 0, \quad \text{with} \quad \delta_{\mathcal{S}} S_T = \int d^d x dt n e^{-\Phi}\text{tr}(\delta K E_{\mathcal{S}}^T). 
\end{equation}
Therefore, we have
\begin{equation}
    E_\mathcal{S}^T = \frac{1}{8\kappa^2}(\mathcal{D}^2\mathcal{S}+ \mathcal{S}(\mathcal{DS})^2- \mathcal{D}\Phi\mathcal{DS}) - \frac{e^{\Phi}}{4}\mathcal{S}\eta \Bar{\mathcal{T}} = 0, \label{eq:EOM_S}
\end{equation}
where we defined the O$(d,d)$ covariant energy-momentum tensor 
\begin{equation}
    \mathcal{\Bar{T}} \equiv \frac{1}{n}\left(\eta\frac{\delta S_m}{\delta \mathcal{S}} \mathcal{S}- \eta \mathcal{S}\frac{\delta S_m}{\delta \mathcal{S}}\right).
\end{equation}
Notice that the variation of the matter action in (\ref{variationS}) is already unconstrained. This follows from the fact that $\delta \mathcal{S} = - \mathcal{S} \delta \mathcal{S} \mathcal{S}$ which implies
\begin{equation}\label{matteractionproperty}
    \frac{\delta S_m}{\delta \mathcal{S}} = - \mathcal{S}\frac{\delta S_m}{\delta \mathcal{S}}\mathcal{S},
\end{equation}
and it is straightforward to show that this guarantees that 
\begin{equation}
    \text{tr}\left(\delta \mathcal{S} \frac{\delta S_m}{\delta \mathcal{S}}\right) = \text{tr}\left(\delta K \frac{\delta S_m}{\delta \mathcal{S}}\right),
\end{equation}
and so we can write the O$(d,d)$ energy-momentum tensor as 
\begin{equation}
    \mathcal{\Bar{T}} =  -\frac{2}{n}\eta \mathcal{S}\left( \frac{\delta S_m}{\delta \mathcal{S}}\right).
\end{equation}

Varying with respect to $n$ gives
\begin{subequations}
    \begin{equation}
         \delta_n S_T = \int d^d x dt n e^{-\Phi}\frac{\delta n}{n}E_n^T
    \end{equation}
    \begin{equation}
        E_n^T \equiv -\frac{1}{2\kappa^2}\left(-(\mathcal{D}\Phi)^2- \frac{1}{8}\text{tr}(\mathcal{DS})^2\right)+ e^{\Phi}\frac{\delta S_m}{\delta n}.
    \end{equation}
\end{subequations}
Thus, the equation of motion for $n$, $E_n^T =0$, is
\begin{equation}
    (\mathcal{D}\Phi)^2+ \frac{1}{8}\text{tr}(\mathcal{DS})^2 = -2\kappa^2 e^{\Phi}\frac{\delta S_m}{\delta n} = 2\kappa^2 e^{\Phi}\bar{\rho}, \label{eq:EOM_n}
\end{equation}
where in the last equality we considered
\begin{align}
    T_{\mu\nu} = -\frac{2}{\sqrt{-G}}\frac{\delta S_m}{\delta g^{\mu\nu}} &\implies T_{00} 
    = -\frac{n^2}{\sqrt{g}}\frac{\delta S_m}{\delta n},\\
    \text{giving} \quad \frac{\delta S_m}{\delta n} &= - \sqrt{g} \rho\equiv -\bar{\rho}.
\end{align} 
 
As usual in cosmology, rather than working directly with the equations of motion for the matter fields $\chi$, we can use the continuity equation due to the Bianchi identities. Taking time derivative of (\ref{eq:EOM_n}) and using equations (\ref{eq:EOM_S}) and (\ref{eq:EOM_Phi}), we find
\begin{equation}
    \mathcal{D}{\Bar{\rho}} + \frac{1}{4}\text{tr}(\mathcal{S (D S)}\eta \mathcal{\bar{T}}) - \frac{1}{2}\bar{\sigma}\mathcal{D}{\Phi} = 0. \label{eq:covariant_general_continuity}
\end{equation}
Equations (\ref{eq:EOM_Phi}), (\ref{eq:EOM_S}), (\ref{eq:EOM_n}) and (\ref{eq:covariant_general_continuity}) define String Cosmology in a O$(d,d)$ covariant way to lowest order in $\alpha'$ \cite{Gasperini:1991ak}. They reduce to the Pre-Big Bang equations \cite{Gasperini:2002bn} once evaluated in components of the O$(d,d)$ tensors. In the next section we shall find all the duality invariant $\alpha'$-corrections to these equations for a FLRW ansatz.
    
\section{$\alpha'$-corrected action including matter} \label{sec:alpha_corrected_matter}

One might worry that including matter to the corrected action found in \cite{Hohm:2019jgu} would be inconsistent as one would be potentially neglecting $\alpha'$ corrections to the matter sector. In this section, we argue that this is not the case, and that the inclusion of matter should be done given the full set of corrected gravity equations, i.e., that one should consider the following action
\begin{equation}
    S = \frac{1}{2\kappa^2}\int d^d x dt n e^{-\Phi}\left[-(\mathcal{D}\Phi)^2 + \sum_{k=1}^{\infty}\alpha'^{k-1}c_k \text{tr}(\mathcal{DS})^{2k}\right] + S_{m}[\Phi, n, \mathcal{S}, \chi], \label{eq:The_Action_2}
\end{equation}
with $c_1= -1/8$. 

As we show in the following, even if we start with the lowest $\alpha'$ order action coupled with matter (as done in \cite{Gasperini:1991ak}), then include corrections and then perform the field redefinitions as in \cite{Hohm:2019jgu}, the final action has the same functional form as (\ref{eq:The_Action_2}). This follows from the fact that we can use matter field redefinitions to ensure properties $1-5$, as described in Section \ref{sec:lowest_order_vacuum}, regardless of the inclusion of matter as we now show.

We would like to show that the $\ap$-corrected action has the form
\begin{equation}
    S =\frac{1}{2\kappa^2}\int d^dx dt n e^{-\Phi} X(\mathcal{DS}) + S_m[\Phi, n, \mathcal{S}, \chi].
\end{equation}
Suppose that this is true for the $(k-1)$th order in $\alpha'$. Now we consider a field redefinition as in (\ref{eq:field_redef}) such that the action changes as
\begin{align}
    S \rightarrow S_0 &+ \delta S_0 + S_m + \delta S_m + \frac{\alpha'^k}{2\kappa^2} \int d^d x dt n e^{-\Phi} X_k(\{\mathcal{D}\Phi\}, \{\mathcal{S}\}) + \sum_{p=1}^{k-1} S_p +O(\alpha'^{k+1}),
\end{align}
where again the variations of the terms higher order than the $k$th-order are inside O$(\alpha'^{k+1})$. The difference between the present and the previous case is that we have the term $\delta S_m$ that generates corrections to the matter action after the field redefinitions. We have
\begin{equation}
    \delta S_0 + \delta S_m = \delta S_T = \int d^d x dt n e^{-\Phi}\left(\delta \Phi E_{\Phi} + \text{tr}(\delta K E_{\mathcal{S}}) + \frac{\delta n}{n}E_{n}\right).
\end{equation}
Let us check the first property. Assume that the $k$th order action has a term of the form
\begin{equation}
    Z_k = \frac{\alpha'^k}{2\kappa^2} \int d^d x dt n e^{-\Phi} X(\{\mathcal{D}\Phi\}, \{S\})\mathcal{D}^2 \Phi \, .
\end{equation}
Then we can use a field redefinition in $\Phi$ of the form (\ref{eq:field_redef}), which implies
\begin{equation}
    \delta S_T = \alpha'^k \int d^d x dt n e^{-\Phi} \delta \Phi  \left[\frac{1}{2\kappa^2}\left(2\mathcal{D}^2\Phi - (\mathcal{D}\Phi)^2 +\frac{1}{8}\text{tr}(\mathcal{DS})^2\right) + \frac{e^{\Phi}}{n}\frac{\delta S_m}{\delta \Phi}\right],
\end{equation}
such that the $Z_k$ term is cancelled against the first term in $\delta S_T$ by choosing $2\delta \Phi = -X$. Thus, we have 
\begin{equation}
    Z_k + \delta S_T = \frac{\alpha'^k}{2\kappa^2}\int d^d x dt n e^{-\Phi}X \left( \frac{1}{2} (\mathcal{D}\Phi)^2 -\frac{1}{16}\text{tr}(\mathcal{DS})^2 - \kappa^2\frac{e^{\Phi}}{n}\frac{\delta S_m}{\delta \Phi}\right), 
\end{equation}
and everything goes as if we had changed $\mathcal{D}^2\Phi$ in $Z_k$ to
\begin{equation}
    Q'_{\Phi} = \frac{1}{2}(\mathcal{D}\Phi)^2 - \frac{1}{16}\text{tr}(\mathcal{DS})^2 -\kappa^2\frac{e^{\Phi}}{n}\frac{\delta S_m}{\delta \Phi}. \label{eq:property_1_new}
\end{equation}
Comparing with (\ref{eq:property_1}), we see that the last term is the novelty after including matter.

A similar modification also appears to $Q_{\mathcal{S}}$ in (\ref{eq:property_2}). We can get rid of the factor of $\mathcal{D}^2\mathcal{S}$ which appears in
\begin{equation}
    Z_k = \frac{\alpha'^k}{2\kappa^2}\int d^d x dt n e^{-\Phi} X(\{\mathcal{D}\Phi\}, \{\mathcal{DS}\}) \text{tr}(\mathcal{G}\mathcal{D}^2\mathcal{S}), \label{eq:arb_2}
\end{equation}
by a $k$th order field redefinition of $\mathcal{S}$ in (\ref{eq:field_redef}), that would give
\begin{equation}
    \delta S_T = \alpha'^k \int d^d x dt n e^{-\Phi} \text{tr} \left[\delta K\left(\frac{1}{8\kappa^2}(\mathcal{D}^2\mathcal{S}+ \mathcal{S}(\mathcal{DS})^2- \mathcal{D}\Phi\mathcal{DS}) - \frac{e^{\Phi}}{4}\mathcal{S}\eta \Bar{\mathcal{T}}\right) \right]. \label{eq:arb_3}
\end{equation}
If $\delta K/4 = -X \mathcal{G}$, then the first term in (\ref{eq:arb_3}) cancels (\ref{eq:arb_2}) and we get
\begin{equation}
    \delta S_T + Z_k = \frac{\alpha'^k}{2\kappa^2}\int d^d x dt n e^{-\Phi} (-X)\text{tr}\left\{\mathcal{G}\left[\mathcal{S}(\mathcal{DS})^2 - (\mathcal{D}\Phi)(\mathcal{DS}) - 2\kappa^2e^{\Phi}\mathcal{S}\eta \Bar{\mathcal{T}}\right]\right\}.
\end{equation}
The net effect is to change the $\mathcal{D}^2\mathcal{S}$ factor by 
\begin{align}
    Q'_{\mathcal{S}} &= -\mathcal{S}(\mathcal{DS})^2 + (\mathcal{D}\Phi)(\mathcal{DS}) + 2\kappa^2 e^{\Phi}\mathcal{S}\eta \Bar{\mathcal{T}} \label{eq:property_2_new},
\end{align}
and we see that the difference to $Q_{\mathcal{S}}$ in (\ref{eq:property_2}) is only the last term coming from the variation of the matter action.

Note that when considering the properties $1-5$ listed in Section \ref{sec:lowest_order_vacuum}, we would have modifications due to the new $Q'_{\Phi}$ and $Q'_{\mathcal{S}}$ that get modified as compared with the previous $Q_{\Phi}$ and $Q_{\mathcal{S}}$. But by including matter we are also increasing the number of fields that can be used to do field redefinitions: now we can also redefine the matter fields. In fact, the last terms in (\ref{eq:property_1_new}) and (\ref{eq:property_2_new}) can be eliminated by redefinitions of the following schematic form
\begin{equation}
    \chi \rightarrow \chi + \alpha'^k \delta \chi. \label{eq:field_redef_matter}
\end{equation}
Under such a change, the matter action varies as
\begin{equation}
    S_m \rightarrow S_m + \delta S_m, \qquad \delta S_m = \alpha'^k \int d^d x dt n e^{-\Phi} \delta \chi \left(\frac{e^{\Phi}}{n}\frac{\delta S_m}{\delta \chi}\right) \, .
\end{equation}
Thus, if the change of variables is such that
\begin{equation}
    \delta \chi \frac{\delta S_m}{\delta \chi} = \frac{1}{2}X\frac{\delta S_m}{\delta \Phi}, \label{eq:arb_4}
\end{equation}
the last term in (\ref{eq:property_1_new}) cancels with $\delta S_m$ above, and we then have
\begin{equation}
    Q'_{\Phi} \rightarrow Q_{\Phi}.
\end{equation}
Similarly, if the new matter variable is such that 
\begin{equation}
    \delta \chi \frac{\delta S_m}{\delta \chi} = -n X \text{tr}[\mathcal{GS}\eta \Bar{\mathcal{T}}] = -n X\text{tr} \left\{\mathcal{GS}\eta \left[\frac{1}{n}\left(\eta\frac{\delta S_m}{\delta \mathcal{S}} \mathcal{S}- \eta \mathcal{S}\frac{\delta S_m}{\delta \mathcal{S}}\right)\right]\right\}, \label{eq:arb_5}
\end{equation}
then
\begin{equation}
    Q'_{\mathcal{S}} \rightarrow Q_{S}.
\end{equation}
Hence, by using matter field redefinitions of the form (\ref{eq:field_redef_matter}) we can preserve the form of the $Q$'s in the presence of matter coupling, in such a away that the proofs of the properties $1-4$ in \cite{Hohm:2019jgu} still apply. For property 5, we need to consider field redefinitions in the lapse function. Indeed, a term of the form
\begin{equation}
    \frac{\alpha'^k}{2\kappa^2} \int d^d x dt n e^{-\Phi}X(\mathcal{DS})\text{tr}(\mathcal{DS})^2,
\end{equation}
can be cancelled by a term in the variation of $S_T$ under
\begin{equation}
    n \rightarrow n + \alpha'^k \delta n,
\end{equation}
that is,
\begin{equation}
    \delta S_T = \alpha'^k \int d^d x dt n e^{-\Phi}\frac{\delta n}{n}\left[-\frac{1}{2\kappa^2}\left(-(\mathcal{D}\Phi)^2- \frac{1}{8}\text{tr}(\mathcal{DS})^2\right)+ e^{\Phi}\frac{\delta S_m}{\delta n}\right].
\end{equation}
The cancellation is ensured if we choose
\begin{equation}
    \frac{\delta n}{n} = -8 X. 
\end{equation}
That, however, implies that there is an extra term in the matter action,
\begin{equation}
    S_m \rightarrow S_m - \alpha'^k \int d^d x dt n e^{-\Phi} 8 X \left(e^{\Phi}\frac{\delta S_m}{\delta n}\right).
\end{equation}
Fortunately, we can still do a field redefinition on the matter field to also cancel this extra matter action. Indeed, choosing
\begin{equation}
    \delta \chi \frac{\delta S_m}{\delta \chi} = 8 n X \frac{\delta S_m}{\delta n},
\end{equation}
we recover the original form of the matter action.

In order to exemplify how the change of variables in the $\chi$ are explicitly done, let us consider the case that the matter action acts like current terms of the background fields $\Phi$ and $\mathcal{S}$
\begin{equation}
    S_m = \int d^d xdt n e^{-\Phi}\left(J_{\Phi}\Phi + \text{tr}(J_{\mathcal{S}}S)\right).
\end{equation}
Note that all the dependence on the matter fields, that have been denoted schematically $\chi$, is now inside the sources $J$. Thus, a variation in the matter fields should yield a variation in the currents. With that in mind, the conditions (\ref{eq:arb_4}) and (\ref{eq:arb_5}) for preserving the form of the $Q$'s after including the matter action are equivalent to
\begin{equation}
    \delta J_{\Phi} \Phi = \frac{1}{2}X J_{\Phi}, \qquad \text{tr}(\delta J_{\mathcal{S}}S) = 2X \text{tr}\left[\mathcal{G}(J_{\mathcal{S}} - \mathcal{S} J_{\mathcal{S}} \mathcal{S})\right],
\end{equation}
which are explicitly solved by choosing
\begin{equation}
    \delta J_{\Phi} = \frac{X J_{\Phi}}{2 \Phi}, \qquad \delta J_{\mathcal{S}} = 2X \mathcal{G}( J_{\mathcal{S}}\mathcal{S} - \mathcal{S} J_{\mathcal{S}} ).
\end{equation}

Therefore, we have shown that after redefining the matter fields, the inclusion of matter does not invalidate the properties $1-5$ of Section 2. Thus, we can immediately use the results of Section \ref{sec:lowest_order_vacuum} and write the corrections to the $0$th order action (\ref{eq:AlphaCosmo_matter}) as
\begin{equation}
    S_k = \alpha'^k \int d^d x dt n e^{-\Phi}X(\mathcal{DS}),
\end{equation}
where $X(\mathcal{DS})$ only contains first order time derivatives of $\mathcal{S}$ and does not have $\text{tr}(\mathcal{DS})^2$ factors. Finally, the $\ap$-corrected gravitional and matter system is described by
\begin{align}
    S &= -\frac{1}{2\kappa^2}\int d^d x dt n e^{-\Phi}\left[(\mathcal{D}\Phi)^2 + \frac{1}{8}\text{tr}(\mathcal{DS})^2\right] + \sum_{k=1}^{\infty}\frac{\alpha'^k}{2\kappa^2} \int d^d x dt n e^{-\Phi}X_k(\mathcal{DS}) \nonumber\\
    &+ S_m[\Phi, n,\mathcal{S}, \chi].  \label{eq:The_Action!}
\end{align}
In summary, assuming again a FRLW ansatz, the $\alpha'$-corrected action including matter sources has the form (\ref{eq:The_Action_2}) regardless of whether we couple matter before or after field redefinitions. 

\section{$\alpha'$-corrected cosmological equations}\label{sec:alpha_cosmo}

The equations of motion which follow from the action (\ref{eq:The_Action_2}) are
\begin{subequations}\label{eq:EOM_matter}
    \begin{equation}\label{eq:EOM_Phi_matter}
         2\mathcal{D}^2{\Phi} - (\mathcal{D}\Phi)^2 - \sum_{k=1}^{\infty}\alpha'^{k-1}c_k \text{tr}(\mathcal{DS})^{2k} = \kappa^2 e^{\Phi} \bar{\sigma},
    \end{equation}
    \begin{equation}\label{eq:EOM_n_matter}
        (\mathcal{D}\Phi)^2 - \sum_{k=1}^{\infty}\alpha'^{k-1}(2k-1)c_k \text{tr}(\mathcal{DS})^{2k} = 2\kappa^2 \Bar{\rho}e^{\Phi},
    \end{equation}
    \begin{equation}\label{eq:EOM_S_matter}
         \mathcal{D}\left(e^{-\Phi}\sum_{k=1}^{\infty}\alpha'^{k-1}4kc_k\mathcal{S}(\mathcal{DS})^{2k-1}\right) = -\kappa^2 \eta \Bar{\mathcal{T}},
    \end{equation}
\end{subequations}
which can be obtained by varying the action: the variation of the gravitational sector follows as in \cite{Hohm:2019jgu} while the contributions from variations of the matter action were calculated in Section \ref{sec:lowest_order_vacuum}. A consistency check is to note that these equations reduce to (\ref{eq:EOM_Phi}), (\ref{eq:EOM_n}) and (\ref{eq:EOM_S}) once we put $c_k=0$ for $k>1$ and $c_1= -1/8$. Moreover, turning off the matter sources results in equations (\ref{vacuumequations}), (\ref{vaccumequations2}) and (\ref{vacuumequations3}), as expected. 

Another important consistency check is to confirm that the continuity equation (\ref{eq:covariant_general_continuity}) is still satisfied. This should be so, since $\alpha'$-corrections should not violate diffeomorphism invariance and, therefore, they should preserve the Bianchi identities. In fact, taking time derivative of (\ref{eq:EOM_n_matter}) and using the expression for $\Ddot{\Phi}$ from (\ref{eq:EOM_Phi_matter}), we get
\begin{equation}\label{rhodot}
    2\kappa^2e^{\Phi} \mathcal{D}\bar{\rho}= \sum_{k=1}^{\infty}\alpha'^{k-1}c_k\left(2k \mathcal{D}\Phi\text{tr}(\mathcal{DS})^{2k} - (2k-1)\mathcal{D}\text{tr}(\mathcal{DS})^{2k}\right) - \kappa^2 e^{\Phi}\bar{\sigma}\mathcal{D}\Phi,
\end{equation}
while multiplying (\ref{eq:EOM_S_matter}) by $\mathcal{S}\dot{\mathcal{S}}$ and taking the trace gives
\begin{equation}
    -\kappa^2 e^{\Phi} \text{tr} \mathcal{S}\mathcal{DS}\eta\Bar{\mathcal{T}} = \sum_{k=1}^{\infty}\alpha'^{k-1}4kc_k\left( \mathcal{D}\Phi\text{tr}(\mathcal{DS})^{2k} - (2k-1)\text{tr}\mathcal{D}^2\mathcal{S}(\mathcal{DS})^{2k-1}\right),
\end{equation}
which can be used to simplify (\ref{rhodot}), yielding
\begin{equation}
    \mathcal{D}\Bar{\rho} = -\frac{1}{4}\text{tr}(\mathcal{S(DS)}\eta \bar{\mathcal{T}}) + \frac{1}{2}\mathcal{D}\Phi \bar{\sigma}, \label{eq:general_continuity}
\end{equation} 
as expected. Equations (\ref{eq:EOM_matter}) together with the continuity equation above define manifest O$(d,d)$ covariant String Cosmology to all orders in $\alpha'$. Notice that working in a O$(d,d)$ covariant formulation allows us to find a class of new solutions from any given one. For that, suppose that $\{\Phi_{0}, \mathcal{S}_{0}, \bar{\rho}_{0}, \bar{\mathcal{T}}_{0}, \bar{\sigma}_{0}\}$ is a solution to the equations. Then, using global transformations $\Omega \in \text{O}(d,d)$, we can generate a continuous class of solutions $\{\Phi_\Omega, \mathcal{S}_\Omega, \bar{\rho}_\Omega, \bar{\mathcal{T}}_\Omega, \bar{\sigma}_{\Omega}\}$ where
\begin{equation}\label{Oddtransformations}
    \Phi_\Omega = \Phi_0, \quad \mathcal{S}_\Omega = \Omega^{-1}\mathcal{S}_0\Omega, \quad \bar{\rho}_\Omega = \bar{\rho}_0, \quad \bar{\mathcal{T}}_\Omega = \Omega^\text{T}\bar{\mathcal{T}}_0 \Omega, \quad \bar{\sigma}_{\Omega} = \bar{\sigma}_0,
\end{equation}
the transformation for $\mathcal{S}$ following from its definition, $\mathcal{S} = \eta \mathcal{H}$.

To find the $\alpha'$-corrected Friedmann equations, we write 
\begin{equation}
    \mathcal{S} = \begin{pmatrix}
    0 & g \\
    g^{-1} & 0 
    \end{pmatrix},
\end{equation}
with $g = a^2(t) I$ and $n(t)=1$, where $I$ denotes the $d$-dimensional identity matrix. Thus,
\begin{equation}
    \frac{\delta S_m}{\delta \mathcal{S}} = \begin{pmatrix}
    0 & \frac{\delta S_m}{\delta g^{-1}}\\
    \frac{\delta S_m}{\delta g} & 0 
    \end{pmatrix} = 
    -\frac{\sqrt{-G}}{2} \begin{pmatrix}
    0 & Tg \\
    - g^{-1}T & 0 
    \end{pmatrix},
\end{equation}
where $T$ is the space components of the energy momentum tensor of the matter fields,
\begin{align}
     T_{ij} = T_i^{\;l}g_{lj} = (Tg)_{ij} = - \frac{2}{\sqrt{-G}}\frac{\delta S_m}{\delta g^{ij}} \implies Tg = -\frac{2}{\sqrt{-G}} \frac{\delta S_m}{\delta g^{-1}},
\end{align}
and similarly
\begin{equation}
    g^{-1}T = \frac{2}{\sqrt{-G}} \frac{\delta S_m}{\delta g}.
\end{equation}
It is straightforward to show that
\begin{equation}
    \mathcal{S} \frac{\delta S_m}{\delta \mathcal{S}} \mathcal{S} = -\frac{\sqrt{-G}}{2} \begin{pmatrix}
    0 & -Tg\\
    g^{-1}T & 0 
    \end{pmatrix} = 
    - \frac{\delta S_m}{\delta \mathcal{S}}, 
\end{equation}
such that
\begin{equation}
    \bar{\mathcal{T}} = \frac{1}{n} \eta \mathcal{S}\left(\mathcal{S} \frac{\delta S_m}{\delta \mathcal{S}} \mathcal{S} - \frac{\delta S_m}{\delta \mathcal{S}}\right) = \frac{2}{n}\eta \mathcal{S}\left(\mathcal{S} \frac{\delta S_m}{\delta \mathcal{S}}\mathcal{S}\right),
\end{equation}
as it should be due to (\ref{matteractionproperty}). Now, using 
\begin{equation}
    \eta \mathcal{S} = \begin{pmatrix}
    g^{-1} & 0 \\
    0 & g\end{pmatrix},
\end{equation}
we get
\begin{equation}\label{oddT}
    \bar{\mathcal{T}} = \frac{2}{n}\left(-\frac{\sqrt{-G}}{2}\right)\begin{pmatrix}
    0 & - g^{-1}Tg\\
      T & 0 \end{pmatrix} = 
    \sqrt{g}\begin{pmatrix}
    0 & pI\\
    -pI & 0 \end{pmatrix}.
\end{equation}
Note that we have used $T_i^{\;j} = p \delta_i^{\;j}$, where $p$ is the fluid's pressure. Using (see equation (4.5) in \cite{Hohm:2019jgu})
\begin{equation}
    \dot{\mathcal{S}} = 2H \mathcal{J}, \quad \text{where} \quad \mathcal{J} \equiv \begin{pmatrix}
    0 & g\\
    -g^{-1} & 0
    \end{pmatrix} \quad \text{with} \quad \mathcal{J}^2 =-\mathcal{I}, \label{eq:matrix_properties}
\end{equation}
where $\mathcal{I}$ is the $d\times d$ identity matrix, we have 
\begin{equation}
    \frac{d}{dt}\dot{\mathcal{S}}^{2k-1}- \dot{\Phi}\dot{\mathcal{S}}^{2k-1} + \mathcal{S}\dot{\mathcal{S}}^{2k} = (-1)^{k-1} 2^{2k-1}\left[\left(\frac{d}{dt} - \dot{\Phi}\right)H^{2k-1}\right]\mathcal{J}.
\end{equation}
Then, the equation for $\mathcal{S}$ (\ref{eq:EOM_S_matter}) is
\begin{equation}
    \sum_{k=1}^{\infty}\alpha'^{k-1}4kc_k(-1)^{k-1} 2^{2k-1}\left[\left(\frac{d}{dt} - \dot{\Phi}\right)H^{2k-1}\right]\mathcal{J} = -e^{\Phi}\kappa^2 \mathcal{S}\eta \bar{\mathcal{T}},
\end{equation}
and, defining the function $f(H)$ as in \cite{Hohm:2019jgu},
\begin{equation}
    f(H) = 4d \sum_{k=1}^{\infty}(-\alpha')^{k-1}2^{2k}k c_k H^{2k-1},
\end{equation}
we have
\begin{equation}
    \frac{1}{2d}\left[\left(\frac{d}{dt} - \dot{\Phi}\right)f(H)\right] \mathcal{J} = -e^{\Phi}\kappa^2 \mathcal{S}\eta \Bar{\mathcal{T}}   \implies \left[\left(\frac{d}{dt} - \dot{\Phi}\right)f(H)\right] \mathcal{I}= 2de^{\Phi} \kappa^2 \mathcal{S}\eta \Bar{\mathcal{T}} \mathcal{J}. \label{eq:arb_6}
\end{equation}
The matrix product in the right-hand side is simply
\begin{equation}
    \mathcal{S}\eta \Bar{\mathcal{T}}\mathcal{J} =- \sqrt{g}\begin{pmatrix} \label{eq:arb_7}
    p & 0\\
    0 & p
    \end{pmatrix}.
\end{equation}
Taking the trace of equation (\ref{eq:arb_6}), we find 
\begin{equation}
    \left[\left(\frac{d}{dt} - \dot{\Phi}\right)f(H)\right] = - 2 d a^d e^{\Phi} \kappa^2 p,
\end{equation}
which can be written as, 
\begin{equation}
    \frac{d}{dt}\left(e^{-\Phi}f(H)\right) = - 2 d \kappa^2 \bar{p}.
\end{equation}

Using the expression for the derivative of $\mathcal{S}$ in equation (\ref{eq:matrix_properties}), the equation for the lapse (constraint equation) is
\begin{equation}
    \dot{\Phi}^2 + g(H) = 2\kappa^2 \Bar{\rho}e^{\Phi},   \quad \text{with} \quad g(H) = 2d \sum_{k=1}^{\infty}(-\alpha')^{k-1}2^{2k}(2k-1)c_k H^{2k}.
\end{equation}
Finally, the equation for $\Phi$ is
\begin{equation}
    2 \Ddot{\Phi} - \dot{\Phi}^2 + F(H) = -2\kappa^2 e^{\Phi} \bar{\sigma}, \quad \text{with} \quad F(H) = 2d \sum_{k=1}^{\infty}(-\alpha')^{k-1}c_k 2^{2k} H^{2k}.
\end{equation}
Note that
\begin{equation}
    F'(H) = f(H), \quad F(H) + g(H) = Hf(H) \quad \text{and} \quad g'(H) = H f'(H). \label{eq:function_properties}
\end{equation}

In summary, the equations are 
\begin{subequations}\label{eq:Alpha_Fried_eqs}
    \begin{equation}
     \dot{\Phi}^2 + g(H) = 2\kappa^2 e^{\Phi}\Bar{\rho}   \label{eq:Alpha_Fried_eqs_a}
    \end{equation}
    \begin{equation}
     \frac{d}{dt}f(H) - \dot{\Phi}f(H) = -  2d \kappa^2e^{\Phi} \bar{p}   \label{eq:Alpha_Fried_eqs_b}
    \end{equation}
    \begin{equation}
      2 \Ddot{\Phi} - \dot{\Phi}^2 + F(H) = \kappa^2 e^{\Phi} \bar{\sigma},   \label{eq:Alpha_Fried_eqs_c}
    \end{equation}
    \label{eq:Alpha_Fried_eqs}
\end{subequations}
which imply the generalized continuity equation
\begin{equation}
     \dot{\bar{\rho}} + d H \bar{p} -\frac{1}{2}\dot{\Phi}\bar{\sigma} = 0,
\end{equation}
that can also be obtained by evaluating the second term in (\ref{eq:general_continuity}) using the explicit expression for $\bar{\mathcal{T}}$ in (\ref{oddT}). As expected, these equations are invariant under the scale factor duality transformation $a\rightarrow 1/a$, since under this transformation we have
\begin{equation}
    H\rightarrow-H,\quad \Phi \rightarrow \Phi, \quad f(H) \rightarrow - f(H), \quad g(H) \rightarrow g(H), \quad \bar{\rho}\rightarrow \bar{\rho}, \quad \bar{p} \rightarrow - \bar{p}, \quad \bar{\sigma} \rightarrow \bar{\sigma},
\end{equation}
which relies on the fact that the matter action is duality invariant, which is the case if one considers it to be given by a gas of free strings \cite{Gasperini:1991ak, Tseytlin:1991xk}. This is a remnant of the O$(d,d)$ transformation (\ref{Oddtransformations}) for the FLRW background.

\section{Cosmological Solutions}\label{sec:solutions}

Having the $\ap$-corrected equations of motion for a cosmological background, we can start to consider some typical scenarios. For simplification, we assume no dilaton coupling $\sigma =0$ (though see \cite{Angus:2019bqs,Quintin:2018loc}) and a barotropic equation of state of the form $p=w(t) \rho$. Using relations (\ref{eq:function_properties}) we can write equations (\ref{eq:Alpha_Fried_eqs}) as
\begin{subequations}
    \begin{equation}
         \dot{\Phi}^2 + HF'(H) - F(H) = 2\kappa^2 e^{\Phi} \bar{\rho} \label{eq:Alpha_Fried_no_dil_coup_a}
    \end{equation}
    \begin{equation}
        \dot{H}F''(H) - \dot{\Phi}F'(H) = -2\kappa^2 e^{\Phi} d w \bar{\rho} \label{eq:Alpha_Fried_no_dil_coup_b}
    \end{equation}
    \begin{equation}
         2 \Ddot{\Phi} - \dot{\Phi}^2 + F(H) = 0, \label{eq:Alpha_Fried_no_dil_coup_c}
    \end{equation}
    \label{subeq:Alpha_Fried_no_dil_coup}
\end{subequations}
where 
\begin{equation}
    F(H) = 2d \sum_{k=1}^{\infty}(-\alpha')^{k-1}c_k 2^{2k} H^{2k}, \label{eq:F_function}
\end{equation}
and the standard continuity equation follows
\begin{equation}
    \dot{\bar{\rho}} + d H\bar{p} = 0. \label{eq:stand_continuity}
\end{equation}
    
\subsection{Solutions for a constant dilaton}

It is known that having a rolling dilaton leads to violations of the weak equivalence principle \cite{Taylor:1988nw,Damour:2002nv}, which has been tested to ever increasing precision (see for instance \cite{Touboul:2017grn}). Thus, at least concerning late-time cosmology, it is expected that the dilaton  field is constant\footnote{Dilaton stabilization in the context of string gas cosmology was studied in \cite{Danos:2008pv}.}. Moreover, dilaton stabilization is also fundamental from the perturbation theory point of view, since the dilaton modulates the strength of the string coupling and, therefore, its divergence would prevent us of considering the classical regime given by the tree level contributions. In the context of bosonic supergravity at lowest order in $\ap$, a constant dilaton implies a unique equation of state, corresponding to radiation, $w=1/d$. We now study if this remains true both perturbatively and non-perturbatively.

Assuming a constant dilaton, $\phi = \phi_0$, the shifted dilaton is given by
\begin{equation}
    \dot{\Phi} = -d H, \quad \Ddot{\Phi} = -d\dot{H}. \label{eq:arb_8}
\end{equation}
Plugging this into (\ref{eq:Alpha_Fried_no_dil_coup_c}) and then solving for $\dot{H}$, we find
\begin{equation}
    \dot{H} = \frac{1}{2d}(F(H)- d^2H^2). \label{eq:arb_10}
\end{equation}
Adding equations (\ref{eq:Alpha_Fried_no_dil_coup_a}) and (\ref{eq:Alpha_Fried_no_dil_coup_c}), and using (\ref{eq:arb_10}) together with (\ref{eq:arb_8}), we have
\begin{equation}
    -F(H)+ d^2H^2 +H F'(H) = 2\kappa^2 e^{\Phi} \bar{\rho} = 2\kappa^2 e^{2\phi_0} \rho. \label{eq:arb_9}
\end{equation}
Using (\ref{eq:arb_10}) and (\ref{eq:arb_8}), equation (\ref{eq:Alpha_Fried_no_dil_coup_b}) can be written as
\begin{equation}
    \frac{1}{2d}(F(H)- d^2H^2)F''(H) +dHF'(H) = -2\kappa^2 e^{2\phi_0} d w \rho,
 \end{equation}
and then, from equation (\ref{eq:arb_9}) we have
\begin{align}
    (F'' -2d^2w)(F- d^2H^2) &+ 2d^2H F'(1+w) = 0. \label{eq:arb_11}
\end{align}

Therefore, given an equation of state $w(t)$, we can use (\ref{eq:arb_11}) to find $F(H)$ and then (\ref{eq:arb_10}) to find $H(t)$. Then, using the continuity equation (\ref{eq:stand_continuity}), we can solve for $\rho(t)$. Thus, we have a systematic way to generate solutions given an equation of state (for constant dilaton)\footnote{Note, however, that string theory will determine the coefficients $c_k$, and hence most of the solutions found using this procedure might not be solutions of string theory.}.

Now we can easily see that the only equation of state compatible with the $0$th order expansion is the one corresponding to radiation. Consider only the first term in the expansion of $F(H)$ in equation (\ref{eq:arb_11}), then all the terms will be proportional to $H^2$ and we end up with an algebraic equation for $w$ in terms of $d$, that has $w=1/d$ as solution. 

Let us see how this equation of state gets modified when we include $\alpha^{\prime}$-corrections. For that, let us expand $F(H)$ perturbatively as \footnote{Note that this is a different approach than the one described before; instead of assuming a $w(H(t))$ and finding the $F(H)$ from equation (\ref{eq:arb_11}), we are assuming a form for $F(H)$ and then finding the form of $w$.}
\begin{equation}
    F(H) = - dH^2 + \beta_1 \alpha' H^4 + \ldots \label{eq:F_function_exp}\\
\end{equation}
Then, to first order in $\alpha^{\prime}$, (\ref{eq:arb_11}) gives
\begin{align}
    H^2\left[(d+ d^2)(2d + 2d^2 w)- 4 d(d^2 + d^2w)\right]&+ \nonumber\\
    +\beta_1\alpha'H^4\left[-12(d+d^2) - 2(d+ d^2w) + 8d^2(1+w) \right] + \ldots&= 0. \label{eq:arb_14}
\end{align}
Assuming a constant $w$, the vanishing of the $0$th order terms implies a radiation equation of state, but then the $1$st order terms do not vanish. In fact, not only does it seem that a radiation equation of state is not a solution to the perturbative solution of the equations, but no constant equation of state will be a solution. One should have either a non-perturbative solution or a time dependent $w$. Other than these possibilities, the only case left for constant $w$ is to have also $H=H_0$ constant. In this case we are lead to consider de Sitter solutions, discussed below. 

For now, let us consider a varying equation of state $w(t)$. Its time dependence at first order in $\alpha'$ contributes to the second line in (\ref{eq:arb_14}), so that it can cancel those terms by adjusting the parameter in the expansion of $w(t)$. Such kind of cancellation can also occur for higher orders in $\alpha'$. Let us construct explicitly the perturbative solution up to order $\alpha'^2$. Motivated by the expansion of $F$,
\begin{equation}\label{eq:arb_17}
    F(H) = -d H^2-32 d c_2 \alpha'H^4 +128 dc_3\alpha'^2 H^6 -512 d c_4 \alpha'^3 H^8  + \ldots,
\end{equation}
consider the expansion for $w$ in terms of $H^2$,
\begin{equation}
    w(t) = w_0-32 d  c_2 w_2 \alpha' H^2 +128 d c_3 w_3 \alpha'^2 H^4 -512 d c_4 w_4 \alpha'^3 H^6 + \ldots
\end{equation}
Then, equation (\ref{eq:arb_11}) gives
\begin{align}
    2 (d^2-d^3-d^3 w_0+d^4 w_0) H^2 &- \nonumber\\
    -64 (-7 d^2  c_2-2 d^3  c_2+3 d^3  c_2 w_0-d^4 c_2 w_2+d^5 c_2 w_2) \alpha'H^4 &+ \nonumber \\
   +256 (48 d^2 c_2^2-16 d^2 c_3-9 d^3 c_3+5 d^3 c_3 w_0+24 d^4 c_2^2 w_2 &- \nonumber \\ 
  -d^4 c_3 w_3+d^5 c_3 w_3)\alpha'^2 H^6 +\cdots &= 0 .
\end{align}
From the first line, we have $w_0 = 1/d$. Using this result in the coefficient of the second line and demanding it to vanish, we can solve for $w_2$ and get
\begin{equation}
    w_2 = \frac{2 (2+d)}{(d-1) d^2}.
\end{equation}
Using the values of $w_0$ and $w_2$ in the coefficient of the $\alpha'^2H^6$ term and imposing it to be zero, we have
\begin{equation}
    w_3 = \frac{-48 c_2^2-96 d c_2^2-11 c_3+2 d c_3+9 d^2 c_3}{(d-1)^2 d^2 c_3}.
\end{equation}
Following this procedure, we will obtain $w(H(t))$. 

To find the time dependence of $H$, we consider the following ansatz in equation (\ref{eq:arb_10}),
\begin{equation}
    H(t) = \frac{H_0}{t} + \alpha' \frac{H_1}{t^3} + \alpha'^2 \frac{H_2}{t^5} +\cdots
\end{equation}
(where the coefficients $H_i$ are dimensionless) which implies that its LHS is
\begin{equation}
    \dot{H} = -\frac{H_0}{t^2}-\frac{3 H_1 \alpha }{t^4}-\frac{5 H_2 \alpha ^2}{t^6} + \cdots
\end{equation}
On the other hand, under the expansion (\ref{eq:arb_17}) for $F(H)$, the RHS of (\ref{eq:arb_10}) is
\begin{align}
    \frac{1}{2d}(F(H) &- d^2 H^2) = \frac{1}{2d}\left[\frac{-d H_0^2-d^2 H_0^2}{t^2}-\frac{2 \left(16 d c_2 H_0^4+d H_0 H_1+d^2 H_0 H_1\right)}{t^4}\alpha'+\right. \nonumber\\
   &+ \left.\frac{\left(128 d c_3 H_0^6-128 d c_2 H_0^3 H_1-d H_1^2-d^2 H_1^2-2 d
   H_0 H_2-2 d^2 H_0 H_2\right)}{t^6}\alpha'^2 + \cdots \right],
\end{align}
so that we can match the coefficients of the $\alpha'$-expansion order by order on the two sides of equation (\ref{eq:arb_10}) to get the time dependence of $H(t)$. Doing so, we arrive at
\begin{align}
    H_0 &= \frac{2}{1+d}, \\
    H_1 &= \frac{16 c_2 H_0^4}{-3+H_0+d H_0}= \frac{256 c_2}{(1+d)^4}, \\
    H_2 &= \frac{128 c_3 H_0^6-128 c_2 H_0^3 H_1-H_1^2-d H_1^2}{2 \left(-5+H_0+d H_0\right)}= -\frac{4096 \left(-40 c_2^2+(1+d) c_3\right)}{3 (1+d)^7}.
\end{align}

The time evolution of the energy density can be found by using the continuity equation. We have
\begin{align}
    \ln{\left(\frac{\rho(t)}{\rho(t_0)}\right)} = - d \int_{t_0}^{t} dt H(t)(1+w(t)), 
\end{align}
which can be calculated after using the expressions for $w_0$, $w_2$, $w_3$, $H_0$, $H_1$ and $H_2$,
\begin{align}
    &\frac{\rho(t)}{\rho_0} = \frac{t_0^2}{t^2}\left[1-\frac{128 c_2 (5+d) }{(d-1) (1+d)^3}\frac{\left(t^2+t_0^2\right)}{t^2 t_0^2}\alpha'+\right. \nonumber\\
   & + \left.\left(\frac{2048 (d-1) (1+d) (17+13 d) c_3 -8 c_2^2 (d-1) (49+23 d) }{3 (d-1)^2 (1+d)^6}\frac{t^4}{t^4 t_0^4} \right.\right. \nonumber\\
    &+ \left.\left.\frac{2048 (d-1) (1+d) (17+13 d) c_3 + (d-1)
   (49+23 d)}{3 (d-1)^2 (1+d)^6} \frac{t_0^4}{t^4t_0^4} + \frac{3 (5+d)^2}{3 (d-1)^2 (1+d)^6}\frac{t^2t^2_0}{t^4t_0^4} \right)\alpha'^2\right]  +\cdots 
\end{align}
which is schematically of the form
\begin{equation}
    \frac{\rho(t)}{\rho_0} = \frac{t_0^2}{t^2}\left[1 -\left(\frac{1}{t_0^2} + \frac{1}{t^2}\right)A \alpha'+ \left(B\frac{1}{t_0^4} + C \frac{1}{t^4}+ D \frac{1}{t^2t_0^2}\right)\alpha'^2 \right] + \cdots
\end{equation}

We can continue to higher orders in $\alpha'$, leaving all the coefficients in the expansions for $H(t)$, $w(t)$ and $\rho(t)$ in terms of the $c_k$ constants and the number of spatial dimensions $d$.

Therefore, we see that a constant dilaton implies a specific equation of state, completely determined by the number of dimensions and the string coefficients $\{c_k\}$. Conceptually, this is analogous to what one obtains after considering the lowest order equations, for which the equation of state is determined as a function of the number of dimensions $d$ and the first coefficient, $c_1 = -1/8$. What we observe here is that a constant dilaton is compatible with a unique solution for the matter sector as well, but that has to have a time-dependent equation of state.

\subsection{de Sitter solutions} \label{sec:de_Sitter}

Since the current paradigm of early universe cosmology involves a de Sitter like equation of state during an early phase of inflation \cite{Starobinsky:1980te, Guth:1980zm, Brout:1977ix, Sato:1980yn}, it is of interest to study under which conditions de Sitter solutions can emerge from our setup. As was already shown in \cite{Hohm:2019jgu}, there are choices for the set of coefficients $c_k$ which admit de Sitter solutions in the absence of matter. In the following we show that such solutions can also be constructed in the presence of matter. However, in light of the {\it swampland conjectures} on effective field theories consistent with string theory (see e.g. \cite{Obied:2018sgi,Agrawal:2018own} and \cite{Brennan:2017rbf, Palti:2019pca} for reviews) it is questionable whether the solutions we find are actually consistent with the sets of coefficients $c_k$ which follow from string theory. Nevertheless, since our analysis includes all the $\ap$-corrections, it sheds a new light on the issues. Here, we will consider the possibility of obtaining exact de Sitter solutions both in the String and Einstein frames.

Exact de Sitter solutions have $H=H_0 = \text{constant}$. Thus, the equations of motion (\ref{subeq:Alpha_Fried_no_dil_coup}) become
\begin{subequations}
    \begin{equation}
         \dot{\Phi}^2 + H_0F'(H_0) - F(H_0) = 2\kappa^2 e^{\Phi} \bar{\rho} \label{eq:Alpha_de_sitter_a}
    \end{equation}
    \begin{equation}
        \dot{\Phi}F'(H_0) = 2\kappa^2 e^{\Phi} d w \bar{\rho} \label{eq:Alpha_de_sitter_b}
    \end{equation}
    \begin{equation}
         2 \Ddot{\Phi} - \dot{\Phi}^2 + F(H_0) = 0 \label{eq:Alpha_de_sitter_c}.
    \end{equation}
    \label{subeq:Alpha_de_sitter}
\end{subequations}
Taking the time derivative of (\ref{eq:Alpha_de_sitter_b}), we get (assuming $w\neq0$)
\begin{equation}
    \Ddot{\Phi} = \frac{2\kappa^2 e^{\Phi}dw}{F'(H_0)} \bar{\rho}\left(\dot{\Phi} - dH_0w + \frac{\dot{w}}{w}\right) = \dot{\Phi} \left(\dot{\Phi} -dH_0w +\frac{\dot{w}}{w} \right). \label{eq:de_sitter_1}
\end{equation}
Then, using (\ref{eq:de_sitter_1}) in (\ref{eq:Alpha_de_sitter_c}), we can write
\begin{equation}
    \dot{\Phi}^2 + 2 \dot{\Phi} \left(\frac{\dot{w}}{w} - d H_0 w \right) + F(H_0) = 0, \label{eq:phi_squared_1}
\end{equation}
which implies
\begin{equation}
    \dot{\Phi} = d H_0 w - \frac{\dot{w}}{w} \pm \sqrt{\left(\frac{\dot{w}}{w} - dH_0w \right)^2 - F(H_0)}. \label{eq:de_sitter_2}
    \end{equation}
This gives us a prescription for how the shifted dilaton must be varying so that we have a de Sitter solution in String frame \textit{regardless} of the equation of state.

We still have not considered equation (\ref{eq:Alpha_de_sitter_a}), which can be combined with (\ref{eq:Alpha_de_sitter_b}) and (\ref{eq:Alpha_de_sitter_c}), and then using  (\ref{eq:de_sitter_1}) it yields
\begin{equation}
    \dot{\Phi}^2 + \dot{\Phi} \left( \frac{\dot{w}}{w} - d H_0 w - \frac{F'(H_0)}{2dw}\right) + \frac{H_0}{2} F'(H_0) = 0.\label{eq:phi_squared_2}
\end{equation}
Since now we have two quadratic equations for $\dot{\Phi}$, namely (\ref{eq:phi_squared_1}) and (\ref{eq:phi_squared_2}), which must provide the same solutions, their coefficients have to be the same, giving us two conditions
\begin{subequations}
    \begin{equation}
        \frac{\dot{w}}{w} - dH_0w = - \frac{F'(H_0)}{2dw}\label{eq:condition_sitter_1},
    \end{equation}
    \begin{equation}
        F(H_0) = \frac{H_0}{2} F'(H_0), \quad w\neq0. \label{eq:condition_sitter_2}
    \end{equation}
\end{subequations}
We now look closely to some particular cases.

\subsubsection{Constant equation of state}

For a constant equation of state,  (\ref{eq:de_sitter_2}) implies $\dot{\Phi} = \text{const}$ (and we still need to take care of $w=0$ separately). And from (\ref{eq:de_sitter_1}), we know that
\begin{equation}
    \dot{\Phi} = 0 \quad \text{or} \quad \dot{\Phi} = d H_0 w_0. \label{eq:dil_sol_cons_EOS}
\end{equation}
We analyse each case separately in the following.

\paragraph{The $\dot{\Phi} = 0$ case.} From (\ref{eq:Alpha_de_sitter_c}), $F(H_0) = 0$, while from (\ref{eq:Alpha_de_sitter_b}), $\bar{p} = 0$ (or $w=0$). And (\ref{eq:Alpha_de_sitter_a}) implies
\begin{equation}
    H_0 F'(H_0) = 2 \kappa^2 e^{\Phi_0} \bar{\rho}_0,
\end{equation}
and thus the energy density $\rho$ is either decaying exponentially or it is zero. In the latter case, $\bar{\rho}_0=0$, we then have $F'(H_0)=0$ and we recover the solution discussed in \cite{Hohm:2019jgu}. However, in the former case we have $\bar{\rho}_0 \neq 0$ with pressureless matter (this hints that the Hagedorn phase in string cosmology could induce an exponentially expanding phase in the String frame). In both cases, we have a running dilaton, 
\begin{equation}
     -dH_0 + 2\dot{\phi} = \dot{\Phi} = 0 \implies \dot{\phi} = \frac{dH_0}{2},
\end{equation}
so we do not have a de Sitter solution in the Einstein frame. 

This all makes sense only if we can build a function $F(H_0)$ that satisfies
\begin{equation}
    F(H_0) = 0 \quad \text{and} \quad F'(H_0) = \text{const} = \frac{2\kappa^2 e^{\Phi_0} \bar{\rho}_0}{H_0}.
\end{equation}
Following the same prescription of \cite{Hohm:2019jgu}, a general class of $F(H)$ with possible dS Hubble parameters ${\pm H_0^{(i)}}$ takes the form
\begin{equation}
    F(H) = -d H^2 \left( 1 + \sum_{p=1}^\infty  d_p \alpha'^p H^{2p} \right) \prod_{i=1}^k \left[1 - \left( \frac{H}{H_0^{(i)}} \right)^2 \right], \label{eq:F_function_sol_1}
\end{equation}
 where as in \cite{Hohm:2019jgu} the first factor in parenthesis is an arbitrary series in $H^2$ and  $F'(H_0)$ constrains $\{d_p\}$. Note that the bracketed term in the product is only to the first power here since $F'(H_0) \neq 0$. 

\paragraph{The $\dot{\Phi} = d H_0 w_0$ case.} From (\ref{eq:Alpha_de_sitter_c}), we see that $F(H_0) = d^2 H_0^2 w_0^2$, and from the continuity equation we have $\bar{\rho} = \bar{\rho}_0 e^{-\dot{\Phi} t}.$ Meanwhile, equation (\ref{eq:Alpha_de_sitter_a}) implies
\begin{equation}
    H_0 F'(H_0) = 2 \kappa^2 e^{\Phi_0}\bar{\rho}_0,
\end{equation}
and (\ref{eq:Alpha_de_sitter_b}) gives $\bar{p} = w_0 \bar{\rho}$, a barotropic equation of state. This solution is valid for any constant equation of state. In this case, the dilaton evolves as
\begin{equation}
    \dot{\phi} = \frac{dH_0}{2}(1+w_0). 
\end{equation}
Curiously, for $w_0 = -1$ we will have a de Sitter solution in both Einstein and String frames. This opens room for considering inflationary models in the context of $\ap$-corrected cosmology \cite{Bernardo:2019xxx}. If $w_0 \neq -1$, then we only have de Sitter solution in String frame.

Note that for $w_0 \neq 0$, we can also impose (\ref{eq:condition_sitter_2}), so that
\begin{equation}
    H_0^2 = \frac{k^2 e^{\Phi_0}\bar{\rho}_0}{d^2 w_0^2},
\end{equation}
so that the Hubble constant does not have to correspond to the string scale, allowing more viable cosmological models.

As above, this all makes sense only if we can build a function $F(H_0)$ that satisfies
\begin{equation}
    F(H_0) = d^2 H_0^2 w_0^2 \quad \text{and} \quad F'(H_0) = \text{const} = \frac{2\kappa^2 e^{\Phi_0} \bar{\rho}_0}{H_0}. \label{eq:sol_const_w_linear_sdil}
\end{equation}
Naturally, we can use the same prescription as in (\ref{eq:F_function_sol_1}) and write 
\begin{equation}
     F(H) = d^2 H^2 w_0^2 - d H^2 \left( 1 + \sum_{p=1}^\infty f_p \ap^p H^{2p} \right) \prod_{i=1}^k \left[1 - \left( \frac{H}{H_0^{(i)}} \right)^2 \right],\label{eq:F_function_sol_2}
\end{equation}
where $\{f_p\}$ are constrained by the value of $F'(H_0)$. 

\paragraph{The $w=0$ case.} From (\ref{eq:Alpha_de_sitter_b}), we see that $\dot{\Phi}=0$ or $F'(H_0)=0$. The former case was already considered above. For the latter, we can solve (\ref{eq:Alpha_de_sitter_c}) for $\Phi (t)$,
\begin{equation}
    \Phi (t) = \Phi_0 - 2 \ln{\left\{\cosh{\left[\frac{\sqrt{F(H_0)}}{2}(t-t_0)\right]}\right\}}, \label{eq:s_dil_solution_de_sitter}
\end{equation}
where it was assumed $F(H_0)>0$; for $F(H_0)<0$, then we have a cosine inside the log instead. Plugging this solution into (\ref{eq:Alpha_de_sitter_a}), we arrive at
\begin{equation}
    -F(H_0) = 2 \kappa^2 e^{\Phi_0} \bar{\rho}_0.
\end{equation}
Thus, if we do not want to have negative energy density, we need $F(H_0)=0$ and we recover the vacuum case discussed in \cite{Hohm:2019jgu}. The other possibility is to have $F(H_0)<0$, then the energy density decays exponentially with time.

\subsubsection{Constant dilaton}

For $\dot{\phi}=0$, then we have $\dot{\Phi} = -d H_0$ and $\Ddot{\Phi} = 0$. Thus, the equations of motion become
\begin{subequations}
    \begin{equation}
         d^2 H_0^2 + H_0F'(H_0) - F(H_0) = 2 k \rho \label{eq:Alpha_de_sitter_const_dil_a}
    \end{equation}
    \begin{equation}
       H_0 F'(H_0) = -2 k p \label{eq:Alpha_de_sitter_const_dil_b}
    \end{equation}
    \begin{equation}
        - d^2 H_0^2 + F(H_0) = 0 \label{eq:Alpha_de_sitter_const_dil_c},
    \end{equation}
    \label{subeq:Alpha_de_sitter}
\end{subequations}
where $k \equiv \kappa^2 e^{2\phi_0}$. We see right away that 
\begin{equation}
  F(H_0) = d^2 H_0^2 \quad \text{and} \quad H_0 F'(H_0) = -2 k p ,
\end{equation}
which is the same as having $w_0=-1$, corresponding to the same solution we have obtained above in (\ref{eq:sol_const_w_linear_sdil}). We also know from the continuity equation that the energy density and pressure will be constant in this case. 

We can now consider a consistency check: let us impose first that $w = -1$ and $\dot{\phi}=0$ in (\ref{eq:arb_11}), and see if that implies necessarily that $H$ is constant. It is easy to see that the equations imply
\begin{equation}
    (d^2 H^2 - F(H)) ( F''(H) + 2d^2) = 0,
\end{equation}
providing a unique solution $F(H) = d^2 H^2$, which implies $\dot{H}=0$ after (\ref{eq:arb_10}). 
Therefore, the $\ap$-corrected cosmology allows a de Sitter solution which does not violate the weak equivalence principle in the presence of an equation of state $w=-1$ both in the Einstein and String frames.

\subsubsection{Dynamical equation of state and dilaton}

\begin{figure}[h]
    \centering
    \includegraphics[scale=.6]{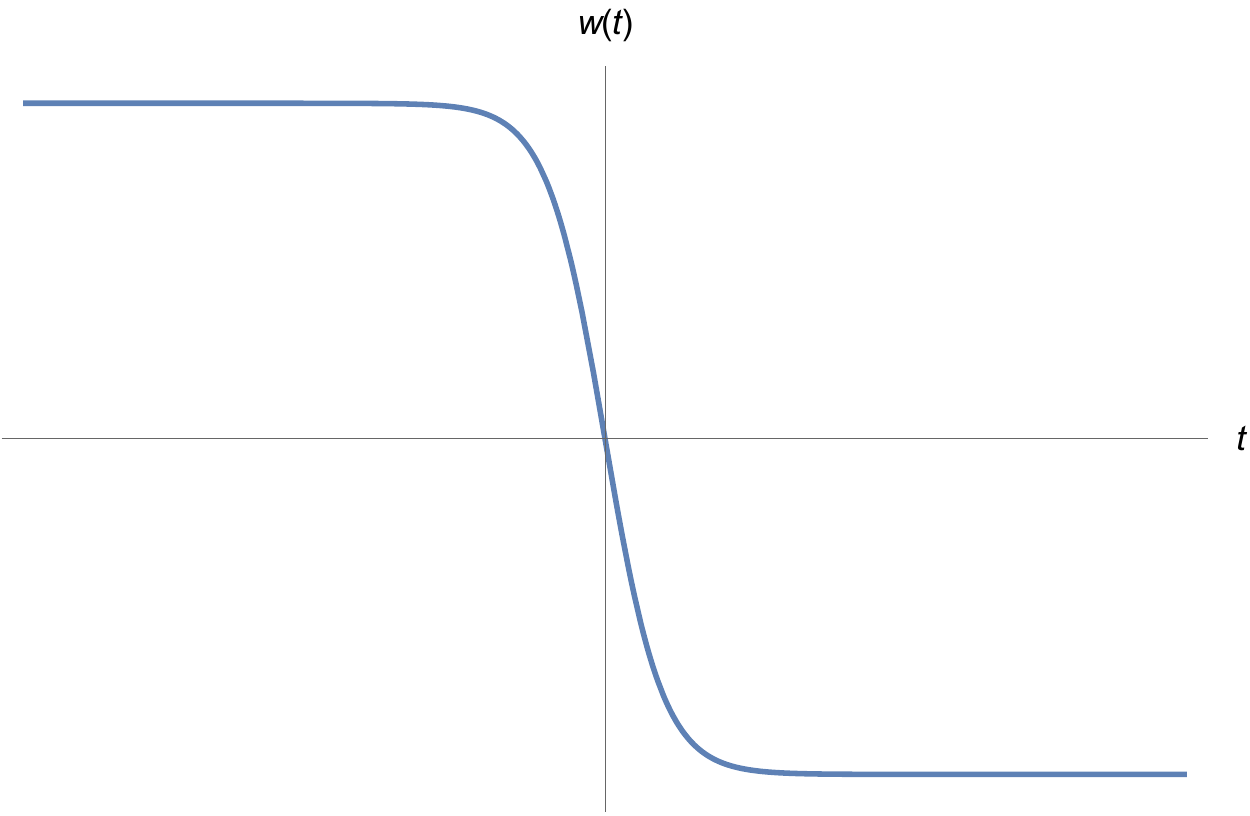}
    \caption{The equation of state is an odd function of time, where the origin is precisely when the shifted dilaton is proportional to the dilaton, i.e., representing the self-dual point of the scale-factor duality.}
    \label{fig:EOS_de_Sitter}
\end{figure}

For the more general case, we can use the solution for the shifted dilaton (\ref{eq:s_dil_solution_de_sitter}). Focusing on having $F(H_0)>0$, we can solve exactly for the energy density and pressure,
\begin{subequations}
    \begin{equation}
        \bar{\rho} (t) = -\frac{e^{-\Phi_0}}{2\kappa^2} \left\{F(H_0) -F'(H_0) H_0 \cosh ^2\left[\frac{1}{2} \sqrt{F(H_0)} (t-t_0)\right]\right\}
    \end{equation}
    \begin{equation}
        \bar{p}(t) = - \frac{e^{-\Phi_0}  F'(H_0)  \sqrt{F(H_0) }}{4 d \kappa^2} \sinh \left[\sqrt{F(H_0) } (t-t_0)\right].
    \end{equation}
\end{subequations}
We see that in order to have the energy density always positive, we need to have
\begin{equation}
    F'(H_0) H_0 > F(H_0), \label{eq:energy_condition}
\end{equation}
which also tells us that $F'(H_0)$ and $H_0$ must have the same sign. The equation of state is
\begin{equation}
    w(t) = \frac{F'(H_0) \sqrt{F(H_0) } \sinh \left[\sqrt{F(H_0) } (t-t_0)\right]}{2 d F(H_0) -2 d F'(H_0) H_0 \cosh ^2\left[\frac{1}{2} \sqrt{F(H_0) } (t-t_0)\right]}.
\end{equation}
Asymptotically we have
\begin{equation}
    w(t\rightarrow \pm \infty) \rightarrow \mp \frac{\sqrt{F(H_0)}}{dH_0}.
\end{equation}
In this limit where the equation of state becomes constant, we can use (\ref{eq:condition_sitter_1}) and (\ref{eq:condition_sitter_2}) to show that $\dot{\Phi} \rightarrow \text{const.}$, recovering exactly the cases given by (\ref{eq:dil_sol_cons_EOS}). A generic plot of the equation of state respecting the energy condition (\ref{eq:energy_condition}) can be seeing in Fig \ref{fig:EOS_de_Sitter}.

\subsubsection{Einstein Frame}

In both scenarios in which we appear to have a quasi-de Sitter expansion in the history of the universe, namely inflation and dark energy, the accelerated expansion is in the Einstein frame. We now look at our equations from its point of view. In \cite{Hohm:2019jgu}, the Hubble parameter in the Einstein frame was shown to be
\begin{equation}
    H'_E (t_E) = - (a(t))^{\frac{d}{d-1}}e^\frac{\Phi}{d-1} \frac{1}{d-1} \left(\dot{\Phi} + H\right),
\end{equation}
where $t_E$ is the cosmic time in the Einstein frame, and it relates to the cosmic time in the String frame, $t$, as
\begin{equation}
    dt = dt_E e^{-\frac{2\phi}{d-1}}, 
\end{equation}
since time has to be reparameterized after assuming $G_{00}=-1$ in String frame. In order to have de Sitter in Einstein frame, we need $dH'_E(t_E)/dt_E=0$, which implies
\begin{equation}
    \Ddot{\Phi} = -\dot{H} - \frac{1}{d-1} (\dot{\Phi} + H) ( d H + \dot{\Phi}). \label{eq:de_Sitter_cond_Einstein}
\end{equation}
Of course we see that if we do not want to have violations of the weak equivalence principle, meaning that we require $\dot{\phi}=0$, then $\dot{\Phi} = -dH$ and the condition implies $\dot{H} = 0$, which is a consistency check at this point. More generally, imposing (\ref{eq:de_Sitter_cond_Einstein}) to (\ref{eq:Alpha_Fried_no_dil_coup_c}), we have
\begin{equation}
       2(d-1) \dot{H} + (d+1) \dot{\Phi}^2  + 2 (d+1) H \dot{\Phi} + 2 d H^2 - (d-1) F(H)= 0.
\end{equation}
For a constant dilaton, it implies 
\begin{equation}
    2\dot{H} = - d^2 H^2 + F(H) = 0,
\end{equation}
since for a constant dilaton de Sitter must be a solution for both frames, recovering the solution $F(H_0) = d^2 H_0^2$.

\section{Conclusions}\label{sec:conclusion}

In this paper we have included matter in an $\Odd$ covariant fashion to the framework developed in \cite{Hohm:2019jgu}, which includes all the $\ap$-corrections that respect this symmetry in a cosmological background. We have shown that matter sources can be either included directly to the $\ap$-corrected equations in vacuum or to the lowest order low-energy effective action for the bosonic supergravity sector and then having it corrected, leading to the same equations of motion. 

Having the non-perturbative equations of motion in $\ap$ for a cosmological background including matter sources, we have considered different cosmological ansatze. Similar to standard results in bosonic supergravity for a vanishing two-form, we have shown that a constant dilaton fixes completely the equation of state perturbatively order by order in the $\ap$-expansion. However, it does not correspond to the radiation equation of state. In fact, perturbatively we have shown that no constant equation of state is possible in such a scenario. One should note that although the equation of state is completely fixed, it depends on the string theory being considered (at the lowest order all string theories have the same coefficient $c_1=-1/8$, which is the only coefficient needed in that case). 

We have also studied de Sitter solutions non-perturbatively, both in Einstein and String frames. We have shown that de Sitter solutions in the String frame are allowed and not restricted to a cosmological constant-like equation of state, as long as the dilaton is evolving (which prevents having the same solution in the Einstein frame). Those solutions have the Hubble constant completely set by the overall energy density instead of the string scale.

We have also shown that for a constant dilaton there is a unique solution for the equation of state that gives de Sitter non-perturbatively in $\ap$: a cosmological constant, $w=-1$. Moreover, this solution has the Hubble constant given by the scale set by overall energy density and not the string scale. This is to be contrasted to the fact that at the lowest order in the $\ap$-expansion the only solution for the equation of state while the dilaton is held constant is the one corresponding to radiation. More generally, we have shown the precise evolution of the equation of state and shifted dilaton so that a de Sitter solutions holds in String frame.

In a future work, we plan to investigate if the $w=-1$ solution is an attractor of the late-time cosmology. In light of the various no-go arguments against de Sitter space in the context of string theory, and the recent constraints on inflationary cosmology \cite{Bedroya:2019tba} resulting from the Trans-Planckian Censorship Conjecture \cite{Bedroya:2019snp}, it is also important to explore the feasibility of other cosmological scenarios (see e.g. \cite{Brandenberger:2011gk} for a comparative review). In particular, it is of interest to explore the possibility of obtaining the quasi-static phase expected from the Hagedorn regime for a hot gas of thermal strings advocated in \cite{Brandenberger:1988aj, Brandenberger:2018wbg}.

\section*{Acknowledgements}

The authors thank Renato Costa for reading the manuscript and relevant discussions. The research at
McGill is supported in part by funds from NSERC and from the Canada Research Chairs program.

\bibliographystyle{JHEP}
\bibliography{HBrefs}

\providecommand{\href}[2]{#2}\begingroup\raggedright\begin{thebibliography}{10}

\bibitem{Polchinski:1998rq}
J.~Polchinski, \emph{{String theory. Vol. 1: An introduction to the bosonic
  string}}, Cambridge Monographs on Mathematical Physics. Cambridge University
  Press, 2007,
  \href{https://doi.org/10.1017/CBO9780511816079}{10.1017/CBO9780511816079}.

\bibitem{Polchinski:1998rr}
J.~Polchinski, \emph{{String theory. Vol. 2: Superstring theory and beyond}},
  Cambridge Monographs on Mathematical Physics. Cambridge University Press,
  2007,
  \href{https://doi.org/10.1017/CBO9780511618123}{10.1017/CBO9780511618123}.

\bibitem{Brandenberger:1988aj}
R.~H. Brandenberger and C.~Vafa, \emph{{Superstrings in the Early Universe}},
  \href{https://doi.org/10.1016/0550-3213(89)90037-0}{\emph{Nucl. Phys.}
  {\bfseries B316} (1989) 391}.

\bibitem{Tseytlin:1991xk}
A.~A. Tseytlin and C.~Vafa, \emph{{Elements of string cosmology}},
  \href{https://doi.org/10.1016/0550-3213(92)90327-8}{\emph{Nucl. Phys.}
  {\bfseries B372} (1992) 443}
  [\href{https://arxiv.org/abs/hep-th/9109048}{{\ttfamily hep-th/9109048}}].

\bibitem{Erdmenger:2009zz}
J.~Erdmenger, ed., \emph{{String cosmology: Modern string theory concepts from
  the Big Bang to cosmic structure}}. 2009.

\bibitem{Baumann:2014nda}
D.~Baumann and L.~McAllister, \emph{{Inflation and String Theory}}, Cambridge
  Monographs on Mathematical Physics. Cambridge University Press, 2015,
  \href{https://doi.org/10.1017/CBO9781316105733}{10.1017/CBO9781316105733},
  [\href{https://arxiv.org/abs/1404.2601}{{\ttfamily 1404.2601}}].

\bibitem{Gasperini:2007zz}
M.~Gasperini, \emph{{Elements of string cosmology}}. Cambridge University
  Press, 2007.

\bibitem{Meissner:1991zj}
K.~A. Meissner and G.~Veneziano, \emph{{Symmetries of cosmological superstring
  vacua}}, \href{https://doi.org/10.1016/0370-2693(91)90520-Z}{\emph{Phys.
  Lett.} {\bfseries B267} (1991) 33}.

\bibitem{Meissner:1991ge}
K.~A. Meissner and G.~Veneziano, \emph{{Manifestly O(d,d) invariant approach to
  space-time dependent string vacua}},
  \href{https://doi.org/10.1142/S0217732391003924}{\emph{Mod. Phys. Lett.}
  {\bfseries A6} (1991) 3397}
  [\href{https://arxiv.org/abs/hep-th/9110004}{{\ttfamily hep-th/9110004}}].

\bibitem{Veneziano:1991ek}
G.~Veneziano, \emph{{Scale factor duality for classical and quantum strings}},
  \href{https://doi.org/10.1016/0370-2693(91)90055-U}{\emph{Phys. Lett.}
  {\bfseries B265} (1991) 287}.

\bibitem{Gasperini:1991ak}
M.~Gasperini and G.~Veneziano, \emph{{$\Odd$ covariant string cosmology}},
  \href{https://doi.org/10.1016/0370-2693(92)90744-O}{\emph{Phys. Lett.}
  {\bfseries B277} (1992) 256}
  [\href{https://arxiv.org/abs/hep-th/9112044}{{\ttfamily hep-th/9112044}}].

\bibitem{Sen:1991zi}
A.~Sen, \emph{{$O(d)\otimes O(d)$ symmetry of the space of cosmological
  solutions in string theory, scale factor duality and two-dimensional black
  holes}}, \href{https://doi.org/10.1016/0370-2693(91)90090-D}{\emph{Phys.
  Lett.} {\bfseries B271} (1991) 295}.

\bibitem{Metsaev:1987zx}
R.~R. Metsaev and A.~A. Tseytlin, \emph{{Order alpha-prime (Two Loop)
  Equivalence of the String Equations of Motion and the Sigma Model Weyl
  Invariance Conditions: Dependence on the Dilaton and the Antisymmetric
  Tensor}}, \href{https://doi.org/10.1016/0550-3213(87)90077-0}{\emph{Nucl.
  Phys.} {\bfseries B293} (1987) 385}.

\bibitem{Hull:1987pc}
C.~M. Hull and P.~K. Townsend, \emph{{The Two Loop Beta Function for $\sigma$
  Models With Torsion}},
  \href{https://doi.org/10.1016/0370-2693(87)91331-1}{\emph{Phys. Lett.}
  {\bfseries B191} (1987) 115}.

\bibitem{Gross:1986mw}
D.~J. Gross and J.~H. Sloan, \emph{{The Quartic Effective Action for the
  Heterotic String}},
  \href{https://doi.org/10.1016/0550-3213(87)90465-2}{\emph{Nucl. Phys.}
  {\bfseries B291} (1987) 41}.

\bibitem{Siegel:1993th}
W.~Siegel, \emph{{Superspace duality in low-energy superstrings}},
  \href{https://doi.org/10.1103/PhysRevD.48.2826}{\emph{Phys. Rev.} {\bfseries
  D48} (1993) 2826} [\href{https://arxiv.org/abs/hep-th/9305073}{{\ttfamily
  hep-th/9305073}}].

\bibitem{Hull:2009mi}
C.~Hull and B.~Zwiebach, \emph{{Double Field Theory}},
  \href{https://doi.org/10.1088/1126-6708/2009/09/099}{\emph{JHEP} {\bfseries
  09} (2009) 099} [\href{https://arxiv.org/abs/0904.4664}{{\ttfamily
  0904.4664}}].

\bibitem{Hohm:2013bwa}
O.~Hohm, D.~Lüst and B.~Zwiebach, \emph{{The Spacetime of Double Field Theory:
  Review, Remarks, and Outlook}},
  \href{https://doi.org/10.1002/prop.201300024}{\emph{Fortsch. Phys.}
  {\bfseries 61} (2013) 926} [\href{https://arxiv.org/abs/1309.2977}{{\ttfamily
  1309.2977}}].

\bibitem{Aldazabal:2013sca}
G.~Aldazabal, D.~Marques and C.~Nunez, \emph{{Double Field Theory: A
  Pedagogical Review}},
  \href{https://doi.org/10.1088/0264-9381/30/16/163001}{\emph{Class. Quant.
  Grav.} {\bfseries 30} (2013) 163001}
  [\href{https://arxiv.org/abs/1305.1907}{{\ttfamily 1305.1907}}].

\bibitem{Hohm:2013jaa}
O.~Hohm, W.~Siegel and B.~Zwiebach, \emph{{Doubled $\alpha'$-geometry}},
  \href{https://doi.org/10.1007/JHEP02(2014)065}{\emph{JHEP} {\bfseries 02}
  (2014) 065} [\href{https://arxiv.org/abs/1306.2970}{{\ttfamily 1306.2970}}].

\bibitem{Hohm:2014xsa}
O.~Hohm and B.~Zwiebach, \emph{{Double field theory at order $\alpha'$}},
  \href{https://doi.org/10.1007/JHEP11(2014)075}{\emph{JHEP} {\bfseries 11}
  (2014) 075} [\href{https://arxiv.org/abs/1407.3803}{{\ttfamily 1407.3803}}].

\bibitem{Marques:2015vua}
D.~Marques and C.~A. Nunez, \emph{{T-duality and $\alpha'$-corrections}},
  \href{https://doi.org/10.1007/JHEP10(2015)084}{\emph{JHEP} {\bfseries 10}
  (2015) 084} [\href{https://arxiv.org/abs/1507.00652}{{\ttfamily
  1507.00652}}].

\bibitem{Baron:2017dvb}
W.~H. Baron, J.~J. Fernandez-Melgarejo, D.~Marques and C.~Nunez, \emph{{The Odd
  story of $\alpha'$-corrections}},
  \href{https://doi.org/10.1007/JHEP04(2017)078}{\emph{JHEP} {\bfseries 04}
  (2017) 078} [\href{https://arxiv.org/abs/1702.05489}{{\ttfamily
  1702.05489}}].

\bibitem{Wu:2013sha}
H.~Wu and H.~Yang, \emph{{Double Field Theory Inspired Cosmology}},
  \href{https://doi.org/10.1088/1475-7516/2014/07/024}{\emph{JCAP} {\bfseries
  1407} (2014) 024} [\href{https://arxiv.org/abs/1307.0159}{{\ttfamily
  1307.0159}}].

\bibitem{Brandenberger:2018bdc}
R.~Brandenberger, R.~Costa, G.~Franzmann and A.~Weltman, \emph{{T-dual
  cosmological solutions in double field theory}},
  \href{https://doi.org/10.1103/PhysRevD.99.023531}{\emph{Phys. Rev.}
  {\bfseries D99} (2019) 023531}
  [\href{https://arxiv.org/abs/1809.03482}{{\ttfamily 1809.03482}}].

\bibitem{Bernardo:2019pnq}
H.~Bernardo, R.~Brandenberger and G.~Franzmann, \emph{{$T$-dual cosmological
  solutions in double field theory. II.}},
  \href{https://doi.org/10.1103/PhysRevD.99.063521}{\emph{Phys. Rev.}
  {\bfseries D99} (2019) 063521}
  [\href{https://arxiv.org/abs/1901.01209}{{\ttfamily 1901.01209}}].

\bibitem{Angus:2018mep}
S.~Angus, K.~Cho and J.-H. Park, \emph{{Einstein Double Field Equations}},
  \href{https://doi.org/10.1140/epjc/s10052-018-5982-y}{\emph{Eur. Phys. J.}
  {\bfseries C78} (2018) 500}
  [\href{https://arxiv.org/abs/1804.00964}{{\ttfamily 1804.00964}}].

\bibitem{Angus:2019bqs}
S.~Angus, K.~Cho, G.~Franzmann, S.~Mukohyama and J.-H. Park,
  \emph{{$\mathbf{O}(D,D)$ completion of the Friedmann equations}},
  \href{https://arxiv.org/abs/1905.03620}{{\ttfamily 1905.03620}}.

\bibitem{Meissner:1996sa}
K.~A. Meissner, \emph{{Symmetries of higher order string gravity actions}},
  \href{https://doi.org/10.1016/S0370-2693(96)01556-0}{\emph{Phys. Lett.}
  {\bfseries B392} (1997) 298}
  [\href{https://arxiv.org/abs/hep-th/9610131}{{\ttfamily hep-th/9610131}}].

\bibitem{Hohm:2019jgu}
O.~Hohm and B.~Zwiebach, \emph{{Duality invariant cosmology to all orders in
  $\alpha$'}}, \href{https://doi.org/10.1103/PhysRevD.100.126011}{\emph{Phys.
  Rev.} {\bfseries D100} (2019) 126011}
  [\href{https://arxiv.org/abs/1905.06963}{{\ttfamily 1905.06963}}].

\bibitem{Brandenberger:2008nx}
R.~H. Brandenberger, \emph{{String Gas Cosmology}},  in \emph{{String
  Cosmology, J.Erdmenger (Editor). Wiley, 2009. p.193-230}}, pp.~193--230,
  2008, \href{https://arxiv.org/abs/0808.0746}{{\ttfamily 0808.0746}}.

\bibitem{Battefeld:2005av}
T.~Battefeld and S.~Watson, \emph{{String gas cosmology}},
  \href{https://doi.org/10.1103/RevModPhys.78.435}{\emph{Rev. Mod. Phys.}
  {\bfseries 78} (2006) 435}
  [\href{https://arxiv.org/abs/hep-th/0510022}{{\ttfamily hep-th/0510022}}].

\bibitem{Gasperini:2002bn}
M.~Gasperini and G.~Veneziano, \emph{{The Pre - big bang scenario in string
  cosmology}}, \href{https://doi.org/10.1016/S0370-1573(02)00389-7}{\emph{Phys.
  Rept.} {\bfseries 373} (2003) 1}
  [\href{https://arxiv.org/abs/hep-th/0207130}{{\ttfamily hep-th/0207130}}].

\bibitem{Quintin:2018loc}
J.~Quintin, R.~H. Brandenberger, M.~Gasperini and G.~Veneziano, \emph{{Stringy
  black-hole gas in $\alpha^{\prime}$-corrected dilaton gravity}},
  \href{https://doi.org/10.1103/PhysRevD.98.103519}{\emph{Phys. Rev.}
  {\bfseries D98} (2018) 103519}
  [\href{https://arxiv.org/abs/1809.01658}{{\ttfamily 1809.01658}}].

\bibitem{Hohm:2019ccp}
O.~Hohm and B.~Zwiebach, \emph{{Non-perturbative de Sitter vacua via $\alpha'$
  corrections}}, \href{https://doi.org/10.1142/S0218271819430028}{\emph{Int. J.
  Mod. Phys.} {\bfseries D28} (2019) 1943002}
  [\href{https://arxiv.org/abs/1905.06583}{{\ttfamily 1905.06583}}].

\bibitem{Brennan:2017rbf}
T.~D. Brennan, F.~Carta and C.~Vafa, \emph{{The String Landscape, the
  Swampland, and the Missing Corner}},
  \href{https://doi.org/10.22323/1.305.0015}{\emph{PoS} {\bfseries TASI2017}
  (2017) 015} [\href{https://arxiv.org/abs/1711.00864}{{\ttfamily
  1711.00864}}].

\bibitem{Palti:2019pca}
E.~Palti, \emph{{The Swampland: Introduction and Review}},
  \href{https://doi.org/10.1002/prop.201900037}{\emph{Fortsch. Phys.}
  {\bfseries 67} (2019) 1900037}
  [\href{https://arxiv.org/abs/1903.06239}{{\ttfamily 1903.06239}}].

\bibitem{Krishnan:2019mkv}
C.~Krishnan, \emph{{de Sitter, $\alpha^{\prime}$-Corrections \& Duality
  Invariant Cosmology}},  \href{https://arxiv.org/abs/1906.09257}{{\ttfamily
  1906.09257}}.

\bibitem{Wang:2019mwi}
P.~Wang, H.~Wu and H.~Yang, \emph{{Are nonperturbative AdS vacua possible in
  bosonic string theory?}},
  \href{https://doi.org/10.1103/PhysRevD.100.046016}{\emph{Phys. Rev.}
  {\bfseries D100} (2019) 046016}
  [\href{https://arxiv.org/abs/1906.09650}{{\ttfamily 1906.09650}}].

\bibitem{Wang:2019kez}
P.~Wang, H.~Wu, H.~Yang and S.~Ying, \emph{{Non-singular string cosmology via
  $\alpha^{\prime}$ corrections}},
  \href{https://doi.org/10.1007/JHEP10(2019)263}{\emph{JHEP} {\bfseries 10}
  (2019) 263} [\href{https://arxiv.org/abs/1909.00830}{{\ttfamily
  1909.00830}}].

\bibitem{Wang:2019dcj}
P.~Wang, H.~Wu, H.~Yang and S.~Ying, \emph{{Construct $\alpha^{\prime}$
  corrected or loop corrected solutions without curvature singularities}},
  \href{https://doi.org/10.1007/JHEP01(2020)164}{\emph{JHEP} {\bfseries 01}
  (2020) 164} [\href{https://arxiv.org/abs/1910.05808}{{\ttfamily
  1910.05808}}].

\bibitem{Taylor:1988nw}
T.~R. Taylor and G.~Veneziano, \emph{{Dilaton Couplings at Large Distances}},
  \href{https://doi.org/10.1016/0370-2693(88)91290-7}{\emph{Phys. Lett.}
  {\bfseries B213} (1988) 450}.

\bibitem{Damour:2002nv}
T.~Damour, F.~Piazza and G.~Veneziano, \emph{{Violations of the equivalence
  principle in a dilaton runaway scenario}},
  \href{https://doi.org/10.1103/PhysRevD.66.046007}{\emph{Phys. Rev.}
  {\bfseries D66} (2002) 046007}
  [\href{https://arxiv.org/abs/hep-th/0205111}{{\ttfamily hep-th/0205111}}].

\bibitem{Touboul:2017grn}
P.~Touboul et~al., \emph{{MICROSCOPE Mission: First Results of a Space Test of
  the Equivalence Principle}},
  \href{https://doi.org/10.1103/PhysRevLett.119.231101}{\emph{Phys. Rev. Lett.}
  {\bfseries 119} (2017) 231101}
  [\href{https://arxiv.org/abs/1712.01176}{{\ttfamily 1712.01176}}].

\bibitem{Danos:2008pv}
R.~J. Danos, A.~R. Frey and R.~H. Brandenberger, \emph{{Stabilizing moduli with
  thermal matter and nonperturbative effects}},
  \href{https://doi.org/10.1103/PhysRevD.77.126009}{\emph{Phys. Rev.}
  {\bfseries D77} (2008) 126009}
  [\href{https://arxiv.org/abs/0802.1557}{{\ttfamily 0802.1557}}].

\bibitem{Starobinsky:1980te}
A.~A. Starobinsky, \emph{{A New Type of Isotropic Cosmological Models Without
  Singularity}},
  \href{https://doi.org/10.1016/0370-2693(80)90670-X}{\emph{Phys. Lett.}
  {\bfseries 91B} (1980) 99}.

\bibitem{Guth:1980zm}
A.~H. Guth, \emph{{The Inflationary Universe: A Possible Solution to the
  Horizon and Flatness Problems}},
  \href{https://doi.org/10.1103/PhysRevD.23.347}{\emph{Phys. Rev.} {\bfseries
  D23} (1981) 347}.

\bibitem{Brout:1977ix}
R.~Brout, F.~Englert and E.~Gunzig, \emph{{The Creation of the Universe as a
  Quantum Phenomenon}},
  \href{https://doi.org/10.1016/0003-4916(78)90176-8}{\emph{Annals Phys.}
  {\bfseries 115} (1978) 78}.

\bibitem{Sato:1980yn}
K.~Sato, \emph{{First Order Phase Transition of a Vacuum and Expansion of the
  Universe}}, {\emph{Mon. Not. Roy. Astron. Soc.} {\bfseries 195} (1981) 467}.

\bibitem{Obied:2018sgi}
G.~Obied, H.~Ooguri, L.~Spodyneiko and C.~Vafa, \emph{{De Sitter Space and the
  Swampland}},  \href{https://arxiv.org/abs/1806.08362}{{\ttfamily
  1806.08362}}.

\bibitem{Agrawal:2018own}
P.~Agrawal, G.~Obied, P.~J. Steinhardt and C.~Vafa, \emph{{On the Cosmological
  Implications of the String Swampland}},
  \href{https://doi.org/10.1016/j.physletb.2018.07.040}{\emph{Phys. Lett.}
  {\bfseries B784} (2018) 271}
  [\href{https://arxiv.org/abs/1806.09718}{{\ttfamily 1806.09718}}].

\bibitem{Bernardo:2019xxx}
H.~Bernardo, R.~Brandenberger and G.~Franzmann, \emph{{\textit{in prep.}}}, .

\bibitem{Bedroya:2019tba}
A.~Bedroya, R.~Brandenberger, M.~Loverde and C.~Vafa, \emph{{Trans-Planckian
  Censorship and Inflationary Cosmology}},
  \href{https://arxiv.org/abs/1909.11106}{{\ttfamily 1909.11106}}.

\bibitem{Bedroya:2019snp}
A.~Bedroya and C.~Vafa, \emph{{Trans-Planckian Censorship and the Swampland}},
  \href{https://arxiv.org/abs/1909.11063}{{\ttfamily 1909.11063}}.

\bibitem{Brandenberger:2011gk}
R.~H. Brandenberger, \emph{{Introduction to Early Universe Cosmology}},
  \href{https://doi.org/10.22323/1.124.0001}{\emph{PoS} {\bfseries ICFI2010}
  (2010) 001} [\href{https://arxiv.org/abs/1103.2271}{{\ttfamily 1103.2271}}].

\bibitem{Brandenberger:2018wbg}
R.~H. Brandenberger, \emph{{Beyond Standard Inflationary Cosmology}},
  \href{https://arxiv.org/abs/1809.04926}{{\ttfamily 1809.04926}}.

\end{thebibliography}\endgroup

\end{document}